\documentclass[preprint,10pt,5p,mathalfa=cal=euler]{elsarticle}

\usepackage{siunitx}
\usepackage{stfloats}
\usepackage{lmodern}
\usepackage[T1]{fontenc}

\journal{}

\begin{document}

\begin{frontmatter}



\title{Machine-Learning Assisted Optimization Strategies for Phase Change Materials Embedded within Electronic Packages}


\author[inst1]{Meghavin Bhatasana}

\affiliation[inst1]{organization={School of Mechanical Engineering and Birck Nanotechnology Center, Purdue University},
            addressline={585 Purdue Mall}, 
            city={West Lafayette},
            postcode={47907}, 
            state={IN},
            country={USA}}

\author[inst1]{Amy Marconnet}

\begin{abstract}
Leveraging the latent heat of phase change materials (PCMs) can reduce the peak temperatures and transient variations in temperature in electronic devices. But as the power levels increase, the thermal conduction pathway from the heat source to the heat sink limits the effectiveness of these systems. In this work, we evaluate embedding the PCM within the silicon device layer of an electronic device to minimize the thermal resistance between the source and the PCM to minimize this thermal resistance and enhance the thermal performance of the device. The geometry and material properties of the embedded PCM regions are optimized using a combination of parametric and machine learning algorithms. For a fixed geometry, considering commercially available materials, Solder 174 significantly outperforms other organic and metallic PCMs. Also with a fixed geometry, the optimal melting points to minimize the peak temperature is higher than the optimal melting point to minimize the amplitude of the transient temperature oscillation, and both optima increase with increasing heater power. Extending beyond conventional optimization strategies, genetic algorithms and particle swarm optimization with and without neural network surrogate models are used to enable optimization of many geometric and material properties. For the test case evaluated, the optimized geometries and properties are similar between all ML-assisted algorithms, but the computational time depends on the technique. Ultimately, the optimized design with embedded phase change materials reduces the maximum temperature rise by 19\% and the fluctuations by up to 88\% compared to devices without PCM.

\end{abstract}



\end{frontmatter}


\section{Introduction}
\label{sec:Intro}
Miniaturization of electronic components has coincided with the rapid advances in technology. This trend has resulted in higher power densities and, thus, drastic increases in heat generation within the components. Prolonged exposure to high operating temperatures often degrades the functionality of the electronic device and has detrimental effects on its long-term reliability.

Phase change materials (PCMs) have been widely investigated to function as a thermal buffer, particularly for components experiencing transient power loads. PCMs absorb some of the heat generated during periods of high power dissipation and can enable longer periods of use before throttling of the processor or shut-off is required to prevent damage. This also provides greater thermal stability as cyclic melting and freezing of the PCM reduces both the range of temperature fluctuations and restricts that fluctuation about its melting point. Most PCM studies have focused on the functionality of PCM laden heat sinks with investigations of different heat sink geometries, orientations, or power levels \cite{Kandasamy2007ApplicationElectronics, Tomizawa2016ExperimentalDevices, Kamkari2018ExperimentalAngles, Arshad2018ExperimentalDiameter, Ganatra2018ExperimentalDevices, Ji2018Non-uniformFins, Kalbasi2019StudiesSinks, Wang2011Three-dimensionalSink}, as well as comparative studies comparing different PCM types \cite{Krishnan2004ThermalSinks} and evaluating the impact of high thermal conductivity inserts to enhance heat conduction \cite{Dmitruk2020AluminumAccumulator, Ruiz2017InvestigationMaterials}. Although these studies demonstrated extensions in high power operating times before the system overheated, numerous materials and interfaces between the PCM-laden heat sink and heat source prevent their effective use as the heat generation rates increase. 

An effective approach to ensure effective usage of the entire volume of PCM is to integrate it at or near the silicon chip level where the heat is generated. Soupremanien \textit{et al.} \cite{Soupremanien2016IntegrationElectronics} investigated the use of PCMs in power electronics by integrating a metallic PCM ($In_{97}Ag_{3}$) within the ceramic for a direct bonded copper substrate. The integration resulted in a reduction of the hotspot temperature by up to 28~\si{\celsius}. Bonner \textit{et al.} demonstrated that directly integrating a metallic PCM ($In_{52}Sn_{42}$) near transistor channels reduced the hotspot temperatures by 21~\si{\celsius} without any interference with the functionality of the device \cite{Bonner2011DieDevices}. Green, Fedorov, and Joshi \cite{Green2012DynamicCoolers} embedded PCMs coupled with a heat spreader in the silicon die near hotspots and saw an extension in device operating times by over 650\%. Shao and colleagues \cite{Shao2014On-chipSprinting} etched wells into the backside of a Si die and filled them with a PCM (Cerrolow 136) demonstrating the feasibility of the embedded cooling approach in enabling computational sprinting. Gurrum \textit{et al.} \cite{Gurrum2002ThermalMaterials} tested a similar approach in high power electronics by etching PCM microchannels within a SiC die. They demonstrated that the PCMs could yield up to 25~\si{\celsius} temperature reductions. These works provide a foundation for embedding phase change materials within the chip for improved heat transfer performance.

All of above works on embedded PCM cooling used metallic PCMs, which are promising candidates for electronics thermal management, in part, due to their high thermal conductivity. Metallic PCMs often have thermal conductivities two orders of magnitude higher than their counterparts while still having comparable volumetric latent heat \cite{Yang2018AdvancesManagement}. Recent work has demonstrated that for the same boundary conditions, metallic PCMs absorb heat $\sim$ 1 to 2 orders of magnitude faster than other classes of PCMs \cite{Shamberger2020ReviewApplications}. One recent study on metallic PCMs in transient, high powered applications found a significant reduction in system operating temperature when compared to polymeric PCMs (upwards of 60~\si{\celsius} at 160~\si{\watt}) \cite{Gonzalez-Nino2018ExperimentalMitigation}. 

Although these studies highlight the promise of metallic PCMs and embedded cooling, there is still a need to optimize both the material properties of the PCM and the embedded heat sink geometry to achieve a balance of high thermal conduction pathways and high heat storage zones. Machine learning aided computations have been widely used in heat transfer to compute complex mechanisms such as heat transfer in boiling \cite{Liu2018Data-drivenResults} or turbulent flow \cite{Kim2020PredictionNetworks}. They have also been prevalent in thermal management applications such as data center cooling where Neural Networks (NNs) have been used to predict real-time temperature and flow profiles \cite{Athavale2018ArtificialCenters,Song2011MultivariateNetworks} and performance metrics \cite{Shrivastava2007DataNetwork,Gao2014MachineOptimization}. Machine learning has also been used in geometric optimization such as designing microchannel heat sink with optimum thermal performance. Shi \textit{et al.} \cite{Shi2019GeometryChannel} used an evolutionary algorithm to reduce the thermal resistance and pumping power of a microchannel heat sink by 28.7\% and 22.9\% respectively, while maintaining the same water mass flow rate. Li \textit{et al.} \cite{Li2020HeatOptimization} similarly optimized a microchannel using a generalized pattern search algorithm and achieved a performance improvement of 63.41\%.

While many works have investigated phase change materials and the use of machine learning to accelerate design, designing embedded PCM cooling solutions with machine learning optimization has not yet been explored. This study introduces machine learning to design and evaluate embedded PCM cooling within the silicon device layer. First, we describe the model used to predict the thermal performance of the system and introduce the machine learning methods. Then we investigate optimization of an embedded cooling solution building up from (i) single point evaluations of commercially-available PCMs to (i) single parameter optimization of the melt temperature (that can be evaluated without machine learning) to (iii-iv) multi-parameter optimization of geometry and material properties. Finally, we explore tradeoffs in the accuracy of the optimized solution versus the computational time required..

\section{Method}

\begin{figure}[t!]
  \includegraphics[width=\columnwidth ]{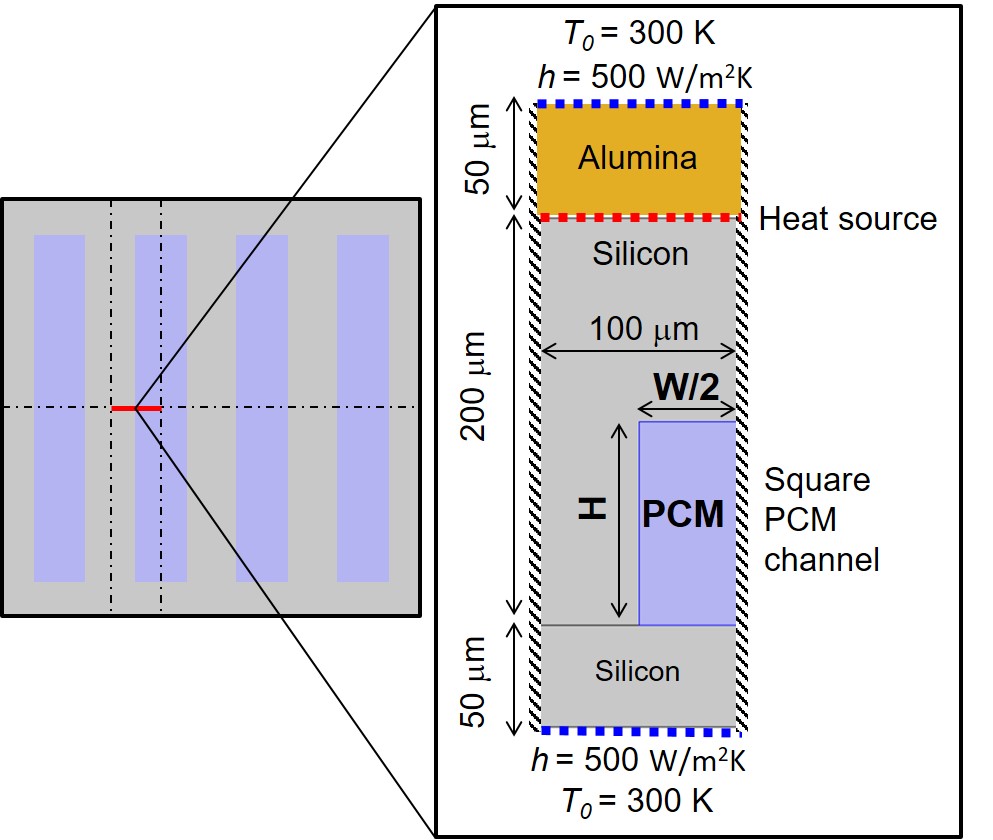}
  \caption{Schematic of the PCM channels (light purple) in the silicon chip (gray). The inset illustrates the 2D cross-section sectioned for analysis based on symmetry (insulated) boundary conditions. Thus the focus is on half of a single channel segment consisting of the silicon, half of a single PCM channel, and an alumina (gold) insulating layer. Heat is generated at the silicon-alumina interface with a square wave pattern with an on time of 0.5~\si{\second}. In these studies, the thermophysical properties of the PCM and the channel height $H$ and width $W$ are varied, while all other thermophysical properties and geometrical parameters are fixed.}
  \label{fig:PCM_Channels}
\end{figure}

Here, we first introduce the system geometry, the computational modelling tool used in this work and the evaluation metrics used to characterize system performance. Then, we describe traditional (\textit{i.e.}, manual optimization) and machine learning aided strategies (\textit{i.e.}, genetic algorithm and particle swarm optimizations with and without neural-network surrogate models) for the optimization studies in this work, along with defining metrics (\textit{i.e.}, overall maximum chip temperature, amplitude of the temperature oscillations, and time to reach a cutoff temperature of 85 \si{\celsius}) that are used for the optimization.

\subsection{Embedded Cooling System Geometry and Boundary Conditions}

Throughout this work, we consider an embedded heat sink consisting of a set of parallel microchannels if height $H$ and width $W$ (see Fig.~\ref{fig:PCM_Channels}) that could be etched in the backside of a 200~\si{\micro \meter} silicon device layer, filled with PCM, and capped with a thin 50~\si{\micro \meter} silicon layer to seal the channels. The top surface of the silicon device layer generates heat and is electrically insulated with a 50~\si{\micro \meter} alumina layer. For the purposes of studying transient heating, the heat generation is a periodic square wave with an on-time of 0.5~\si{\second} of varying amplitude $q''_0$. The top surface of the alumina and the bottom surface of the silicon capping layer are treated as equivalent convection boundary conditions with an effective heat transfer coefficient of $h $ = 500~\si{\watt \per{\square \meter \kelvin}} to represent the thermal pathways to ambient at temperature $T_{0}$ = 300~\si{\kelvin}.

\begin{figure*}[t]
    \centering
    \includegraphics[width=0.85\textwidth]{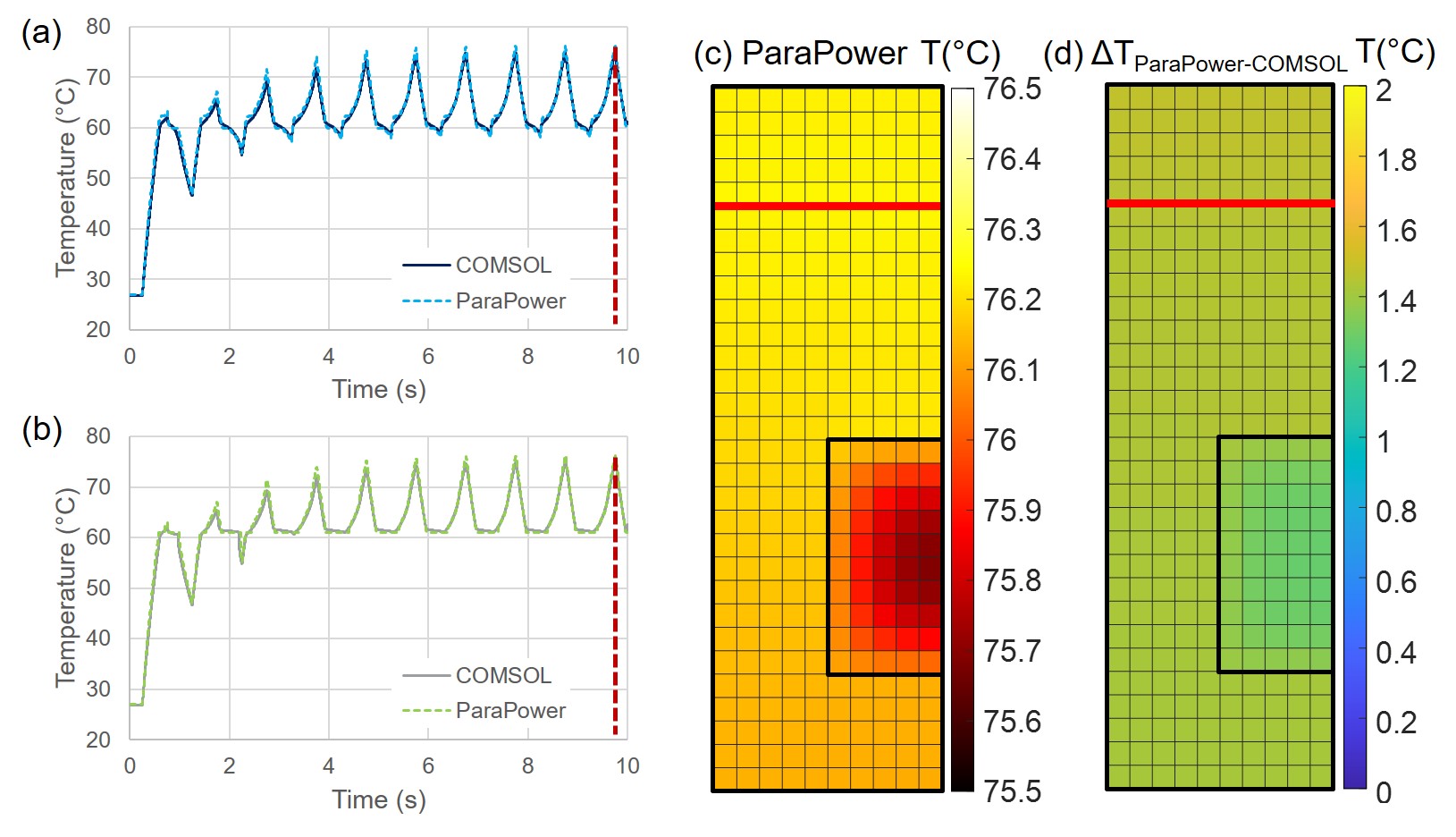}
    \caption{Predicted maximum temperature in (a) the silicon device layer and (b) the PCM for a square wave heat input of $q''_0$ = 75~\si{\kilo \watt \per{\square \meter}} from COMSOL (solid lines) compared to ParaPower (dashed lines) for the channel filled with PureTemp60. (c)Predicted temperature distribution for the ParaPower model at 9.75~\si{\second}, at which temperature in the system is highest (indicated with dashed red line in panels (a-b)). (d) Difference in spatial temperature distribution between ParaPower and COMSOL predictions at 9.75~\si{\second}. ParaPower predictions on average are 1.42~\si{\celsius} higher than COMSOL. The solid red lines in (c-d) indicate the location where heat is generated.  }
    \label{fig:PCMCompareModel}
\end{figure*}

\subsection{Computational Model in ParaPower}

For computational efficiency, thermal analysis is performed using ParaPower, a Matlab-based simulation package developed at the Army Research Lab. Detailed information about ParaPower can be found in \cite{Deckard2019ConvergencePackaging, Border2019IntegratingAnalysis}, so we only briefly describe it here. ParaPower utilizes a resistance network modeling scheme, in which the simulation space is discretized into volumetric elements. Each element is represented by a node in 3D space, with properties (such as specific heat $c_p$, density $\rho$, thermal conductivity $k$) determined by the material and phase within the volume. The properties of the phase change materials used in this work are shown in Table~\ref{tab:PCMs}. It can analyze large, complicated parametric spaces {\textgreater}100 times faster than finite element analysis (FEA) with reasonable accuracy (typically {\textless}2~\si{\celsius}) \cite{BotelerMultiplePackaging}. To confirm accuracy for our system, we analyze the system illustrated in Fig.~\ref{fig:PCM_Channels} in both ParaPower and a commercial FEA solver (COMSOL) for a square-wave heat input with amplitude of $q''_0 =$ 75~\si{\kilo \watt \per{\square \meter}} and on time of 0.5~\si{\second} using PureTemp60 as the embedded PCM (Fig.~\ref{fig:PCMCompareModel}). One key difference between the modeling approaches is that ParaPower assumes a constant fixed point temperature for the PCM while melting, whereas COMSOL has a transition temperature range specified (here, we use 1.75~\si{\celsius}). The temperature evolution and instantaneous spatial temperature profiles are very similar and lend credibility to using the lower fidelity and more computationally efficient ParaPower model. The mean average error in temperature at the hottest time step is 1.42~\si{\celsius} or 2.97\% of the temperature rise. Evaluation of the impact of the specified transition temperature range, mesh element size, and computation time step are provided in the supplemental information.

\begin{figure}[t!]
  \includegraphics[width=\linewidth]{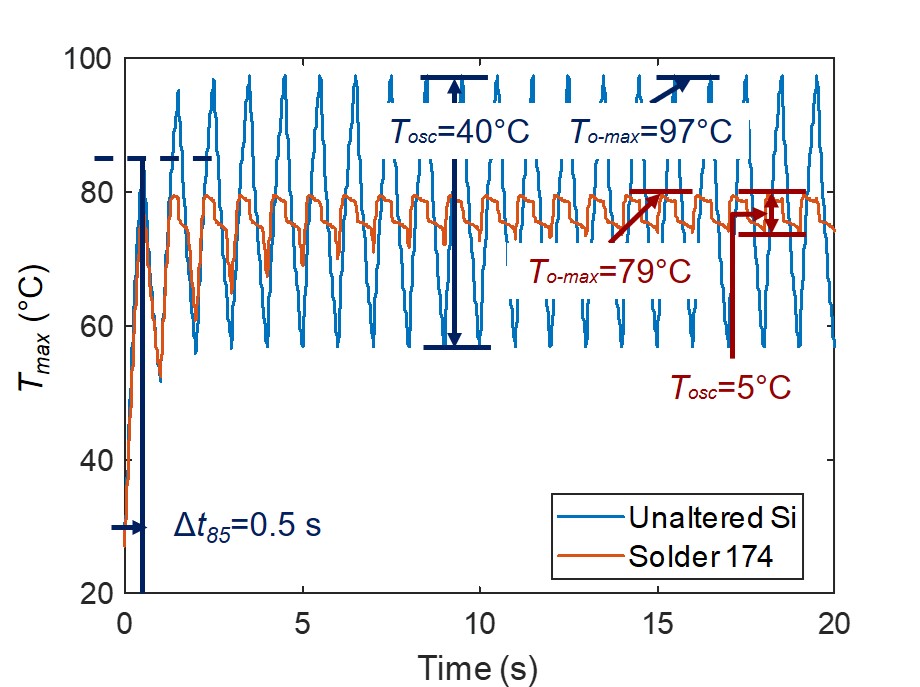}
  \caption{Maximum temperature in the system responding to a square-wave heat input of amplitude $q'' = $100~\si{\kilo \watt \per{\square \meter}} and on time of 0.5~\si{\second} for first 20 seconds of simulation time. Evaluation metrics and their associated values are highlighted for Solder 174 (orange line) compared to the case without PCM (unaltered Si, blue line). Note, although the total simulation time was 1000 seconds (1000 cycles), the system reaches a steady periodic response within 20 seconds (20 cycles). In some cases, longer transient periods are observed and up to 1000 cycles are used to confirm a steady periodic response.}
  \label{fig:Metrics}
\end{figure}

\subsection{Evaluation metrics}

Fig.~\ref{fig:Metrics} illustrates the temporal evolution of maximum temperature in the system for a square wave heat input with amplitude $q''_0 =$ 100~\si{\kilo \watt \per{\square \meter}} with Solder 174 as the embedded PCM compared to device without channels \sloppy (solid silicon). At every time step, the maximum temperature ($T_{o-max}(t)$) is computed. From this, we then define three thermal metrics quantify the effectiveness of embedded PCM cooling from the thermal history:
\begin{enumerate}
\item 	$T_{o-max}$: Overall maximum chip temperature (across all time steps)
\item 	$T_{osc}$: Magnitude of the the oscillations in the transient maximum chip temperature $T_{max}(t)$ at quasi-steady state
\item   $\Delta t_{85}$: Time to a cutoff temperature of 85~\si{\celsius}
\end{enumerate}
These metrics were chosen based on our past work \cite{CollierS.Miers2019THERMALMATERIALS} and discussions with industry collaborators to characterize the relative performance of different materials and systems. To clarify, at a particular instant in time, the transient maximum chip temperature is referenced as $T_{max}$, while $T_{o-max}$ corresponds to the maximum temperature across all simulated time steps.

For the baseline case of a fully silicon chip with no PCM, the $T_{o-max} = 97.3$~\si{\celsius}, $T_{osc} = 40.5$~\si{\celsius}, and $\Delta t_{85} = 0.5$~\si{\second}. Here, Solder 174 significantly outperforms the unaltered Silicon case with a $T_{o-max}$ that is 18~\si{\celsius} lower (or $\approx$ 24\% lower in terms of temperature rise from ambient) and reducing $T_{osc}$ by 88\%. Note that the $T_{o-max}$ does not even exceed the cutoff temperature of 85~\si{\celsius} within the duration of the simulation for Solder 174. 

\begin{table*}[b!]
    \centering
    \caption{Properties of the Commercially-Available PCMs}
    \resizebox{\textwidth}{!}{%
    \begin{tabular}{|c|c|c|c|c|c|c|c|c|c|c|}
        \hline
        \textbf{Number} &
        \textbf{Material} &
          \begin{tabular}[c]{@{}c@{}}$\mathbf{T_m}$ \\ (~\si{\celsius})\end{tabular} &
          \begin{tabular}[c]{@{}c@{}}$\mathbf{\rho_{solid}}$ \\ (\si{\kg \per{\meter \cubed }})\end{tabular} &
          \begin{tabular}[c]{@{}c@{}}$\mathbf{\rho_{liquid}}$ \\ (\si{\kg \per{\meter \cubed }})\end{tabular} &
          \begin{tabular}[c]{@{}c@{}}$\mathbf{k_{solid}}$ \\ (\si{\watt \per{\meter \kelvin}})\end{tabular} &
          \begin{tabular}[c]{@{}c@{}}$\mathbf{k_{liquid}}$ \\ (\si{\watt \per{\meter \kelvin}})\end{tabular} &
          \begin{tabular}[c]{@{}c@{}}$\mathbf{C_{p,solid}}$ \\ (\si{\joule \per{\kilo\gram \kelvin}})\end{tabular} &
          \begin{tabular}[c]{@{}c@{}}$\mathbf{C_{p,liquid}}$ \\ (\si{\joule \per{\kilo\gram \kelvin}})\end{tabular} &
          \begin{tabular}[c]{@{}c@{}}$\mathbf{L_H}$\\ (J/kg)\end{tabular} &
          \textbf{Ref.} 
          \\ \hline
        1 & Cerrolow 117 & 47   & \multicolumn{2}{c|}{9160} & \multicolumn{2}{c|}{15} & 163        & 197        & 36800 & \cite{Shamberger2020ReviewApplications, MatWebIndiumAlloy} 
        \\ \hline
        2 & Cerrolow 136    & 58   & 9060        & 8220       & 33.2       & 10.6      & 323        & 721        & 28900 & \cite{Fukuoka1990NewAlloys} 
        \\ \hline
        3 & Fields Metal & 58.24   & \multicolumn{2}{c|}{7880} & \multicolumn{2}{c|}{19} & \multicolumn{2}{c|}{250} & 31020 & \cite{Gonzalez-Nino2018ExperimentalMitigation, Baez2020MetallicCycles}  
        \\ \hline
        4 & E-BiInSn     & 60.2 & \multicolumn{2}{c|}{8043} & 19.2       & 14.5      & 270        & 297        & 27900 &  \cite{Yang2017ExperimentalFins}
        \\ \hline
        5 & PureTemp60   & 61   & 960        & 870       & 0.25       & 0.15      & 2040        & 2380        & 220000 & \cite{PureTempPureTempSheet} 
        \\ \hline
        6 & Wood's Metal & 70   & \multicolumn{2}{c|}{9670} & 31.6       & 22.4      & 146        & 184        & 40000 & \cite{Shamberger2020ReviewApplications} 
        \\ \hline
        7 & Solder 174   & 77   & 8780        & 8200       & 35.8       & 28.8      & 401        & 883        & 47730 &  \cite{Fukuoka1990NewAlloys} 
        \\ \hline
    \end{tabular}}
    \label{tab:PCMs}
\end{table*}

\subsection{Optimization Strategies \& Studies}

The paper explores four optimization studies with increasing computational complexity:
\begin{enumerate}
  \item Comparing the thermal performance of fixed chip design with seven selected commercial PCMs (see Table~\ref{tab:PCMs}).
  \item Single parameter optimization of the PCM melt temperature for several different input power levels.
  \item	Multi-parameter optimization of the PCM properties for a fixed channel geometry.
  \item	multi-parameter optimization of the channel geometry and the PCM melt temperature (with the other thermophysical properties of the PCM fixed).
\end{enumerate}

While the first two studies can easily be conducted with manual exploration of the entire desired parameter space, the multi-parameter optimization studies would require too much computational time to fully evaluate. Thus, we consider three optimization strategies within this work:
\begin{enumerate}
\item	\textit{Manual optimization} that is an exhaustive parametric search within set boundaries. 
\item	\textit{Direct optimization} where an optimization algorithm is directly tied to ParaPower (\textit{i.e.}, informing the next set of simulation parameters based on previous simulations). 
\item	\textit{Neural network (NN) assisted optimization} that consists of generating a set of training data with ParaPower, using that to train an NN, and then integrating an optimization algorithm to the NN.
\end{enumerate}


\begin{figure}[b!]
  \includegraphics[width=\linewidth]{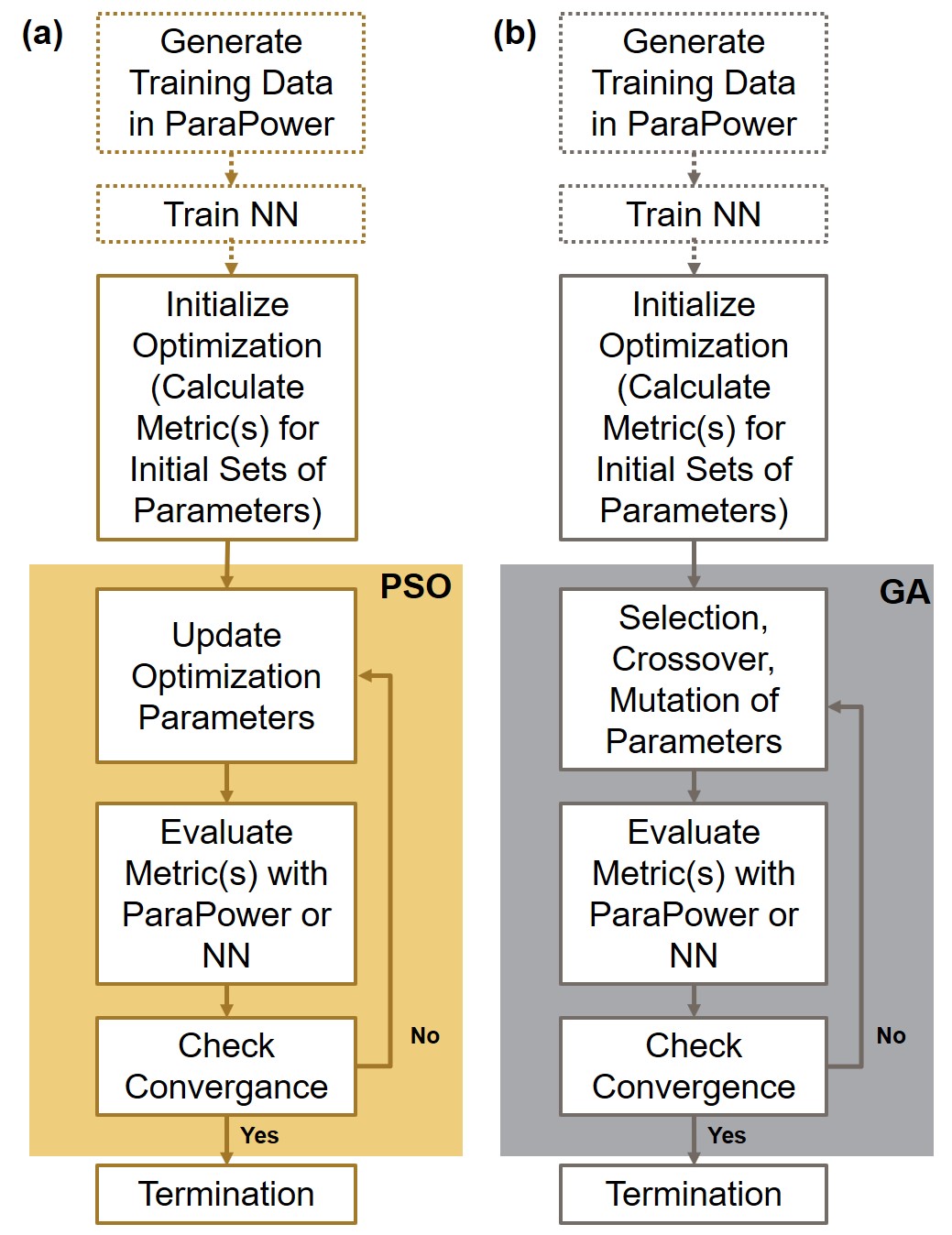}
  \caption{Flowchart detailing the (a) particle swarm optimization and (b) genetic algorithm for optimizing parameters of the system. In both methods, either the the algorithm directly integrates with ParaPower (in which case the algorithm follows just the solid outlined boxes) or first a neural network is trained based on a selected training set of data from ParaPower (illustrated by the the dashed outlined boxes).}
  \label{fig:Algorithms}
\end{figure}
For the second and third optimization strategies, we consider two optimization algorithms: a genetic algorithm (GA) and a particle swarm optimization (PSO). Fig.~\ref{fig:Algorithms} illustrates the different algorithms, with the optional steps of training the neural network. These optimization tools are built into Matlab and described in detail in \cite{MathWorksParticleAlgorithm, MathWorksHowWorks} and thus are only briefly outlined here.

\textit{Genetic algorithms} optimize based on the concept of natural selection that promotes `survival of the fittest'. The algorithm generates a set of initial solutions and calculates the performance metrics for these parameters. The best performing of the set (\textit{i.e.}, the `parents') pass through to the next iteration and produce a new set of solutions (\textit{i.e.}, the `children') by crossing over properties of the parents and introducing randomized mutations. The algorithm terminates when the evaluation metrics for the current set of children do not differ significantly from the past few generations. Thus, the initial set of solutions has now converged or `evolved' to produce the optimum results. 

\textit{Particle swarm optimization} is inspired by the social behavior of birds flocking. Similar to GA, the algorithm generates a set of initial parameters (called `particles' and for example, combinations of thermophyscial properties and geometrical parameters to be optimized) with `velocities' (\textit{i.e.}, directions for parameter changes). For each particle, the evaluation metrics are computed (from ParaPower or the NN) and the best function value of the swarm is calculated. The algorithm subsequently updates the velocity of each particle, evaluates its new position, determines the new best function value from the swarm. If needed, the algorithm updates the best position of the particle throughout its movement history. The velocity of each particle is influenced by its personal best function value in history and the current best function value of the swarm. This process repeats until the algorithm terminates.

 In our work, neural networks function as a surrogate models to enable rapid optimization after being trained with data from ParaPower. NNs are based on the concept of neurons in the brain and are designed to recognize patterns. They are highly robust and have the potential to accurately model any non-linear relationship between independent and dependent variables. They consist of several processing layers that are built of interconnected nodes called neurons. Layers can be broadly classified into three types: input layer, hidden layer, output layer. Processed information that is shared between the neurons is impacted by `weights' that are iteratively adjusted as the NN is trained. The NNs built for this study consist of an input layer that takes in geometric specifications and selected PCM properties, a hidden layer of 10 neurons, and an output layer of a single neuron that either predicts the overall maximum chip temperature $T_{o-max}$ or the oscillations in the maximum chip temperature $T_{osc}$. The network training function for this study was chosen to be Levenberg–Marquardt algorithm since it generally has the fastest convergence and a higher accuracy for function approximation problems \cite{MathWorksChooseFunction}. The non-linear mapping within a NN is based on the ‘activation function’ chosen for the network. When the data from a layer of neurons propagates to the next layer, the neurons of the initial layer are multiplied by their weights, summed together, and passed through the activation function. This introduces non-linearity to an otherwise linear combination of information. The activation function used for the NNs in this study is a Sigmoid function:

\begin{equation}
f(x) = \frac{2}{1 + e^{-2x}} - 1
\label{eq_ASME}
\end{equation}

Thus in this work, we evaluate 5 optimization strategies:
\begin{enumerate}
\item Manual optimization or parametric sweep (labeled PS in figures and tables); 
\item the direct optimization with a genetic algorithm coupled to ParaPower (GA+PP);
\item the direct optimization with particle swarm optimization coupled to ParaPower (PSO+PP);
\item the neural network assisted optimization with a genetic algorithm (GA+NN); and
\item the neural network assisted optimization with particle swarm optimization (PSO+NN).
\end{enumerate}

We generated multiple data sets for the optimization studies using ParaPower to predict performance with different combination of material and geometric parameters. The first data set, consisting of 3300 cases, considered different PCM properties for a fixed channel geometry. This data set was used for the single parameter optimization of the melt temperature (study 2) and the first multi-parameter optimization of properties with a fixed geometry (study 3). The second data set, consisting of 2800 cases, evaluated varying channel geometry (height and width) along with  PCM melt temperatures (with the rest of the PCM properties fixed) and was used for the final multi-parameter optimization of geometry and melting point (study 4). 
To understand the impact of the neural net generation, we repeat the optimization strategies leveraging the neural nets 10 different times. When discussing the results of these NN-assisted approaches, the average of the optimum system property (the PCM properties and/or channel geometric properties), and the ParaPower predicted $T_{o-max}$ and $T_{osc}$ are shown in regular font size. The range of the optimum of the system property and ParaPower predictions of $T_{o-max}$ and $T_{osc}$ are shown as superscript and subscripts respectively.

\section{Results \& Discussion}

\subsection{Comparison of Commercial PCMs in a Fixed Geometry}

\begin{figure*}[b!]
    \includegraphics[width=\textwidth]{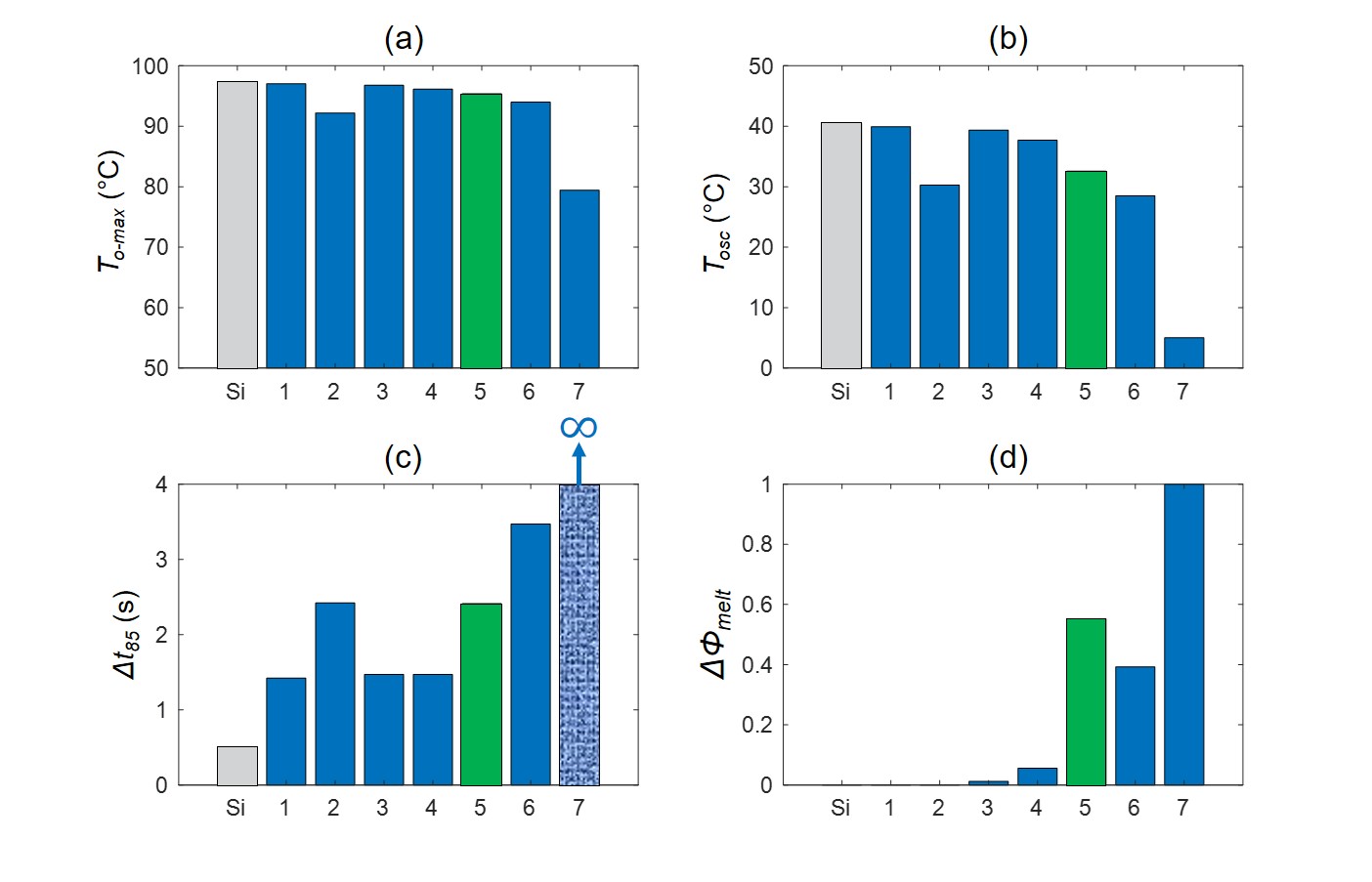}
    \caption{(a) Maximum chip temperature ($T_{o-max}$), (b) chip temperature oscillations ($T_{osc}$), (c) time to 85~\si{\celsius} $\Delta t_{85}$, and (d) oscillation in the PCM melt fraction ($\Delta \Phi_{melt}$) for commercially-available PCMs for a fixed channel geometry with a cyclic heat input of 100~\si{\kilo \watt \per{\square \meter}} compared to a solid silicon chip (silver bar labelled Si). The seven PCMs, described in Table~\ref{tab:PCMs}, are arranged in order of increasing melting temperature. Note that the 7th PCM (Solder 174) never reaches 85~\si{\celsius}.}
    \label{fig:PCMCompare}
\end{figure*}

To first understand the range of performance expected in the system, we considered 6 metallic PCMs and one organic PCM (PureTemp68) that are commercially-available within a fixed geometry. The properties of the PCM are provided in Table~\ref{tab:PCMs}. Note that the melt temperatures range from 47~\si{\celsius} to 77~\si{\celsius}. For this study, the channel size is fixed at 100 ~\si{\micro\meter} $\times$ 50 ~\si{\micro\meter} and Fig.~\ref{fig:PCMCompare} compares the thermal responses via the thermal metrics and the oscillation in the PCM melt fraction for a cyclic heat input with magnitude of 100~\si{\kilo \watt \per{\square \meter}} and on time of 0.5 s. Apart from the silicon, the PCMs are shown in an order of increasing melt temperature and labelled 1- 7 in the graphs corresponding to their labels in Table~\ref{tab:PCMs}.

Approximately, the overall maximum chip temperature \sloppy decreases with increasing melting temperature although it is not monotonic. Solder 174, or case 7, is the best performing material for all 3 thermal metrics. Note that in this case, the system never reaches 85~\si{\celsius} in the duration of the simulation.

Figure~\ref{fig:PCMCompare}(d) shows the oscillation in the percentage of PCM within the channel that is melted (a.k.a., the PCM melt fraction). Note that for the first two PCMs (Cerrolow 117 and Cerrolow 136), the PCM melts completely and remains melted in the quasi-steady region of the curves. Thus, the melt fraction oscillation is zero. Materials 3-6 partially melt and solidify during each cycle. In case 7 (Solder 174), the best performing case, the PCM melt fraction oscillation is 1 indicating that the system fully melts and resolidifes in each cycle. This is the ideal case for the best performance for a given power input and geometry.  

Even though the other PCMs achieve relatively similar overall maximum chip temperatures to the unmodified silicon case, they do perform better than the solid silicon wafer in terms of the short-term performance as illustrated in the longer times to reach the cutoff temperature of 85~\si{\celsius}. Several also perform better in terms of the amplitude of the chip temperature oscillations. In particular, Cerrolow 136 (material 2) shows a 25\% reduction in the chip temperature oscillation. This initial phase of the study shows the importance of optimizing thermophysical properties of the PCM within a fixed geometry and use case. In particular, $T_m$ plays a large role in the performance of the system.

\subsection{Single Parameter Optimization}
 
\begin{figure}[b!]
  \includegraphics[width=\linewidth]{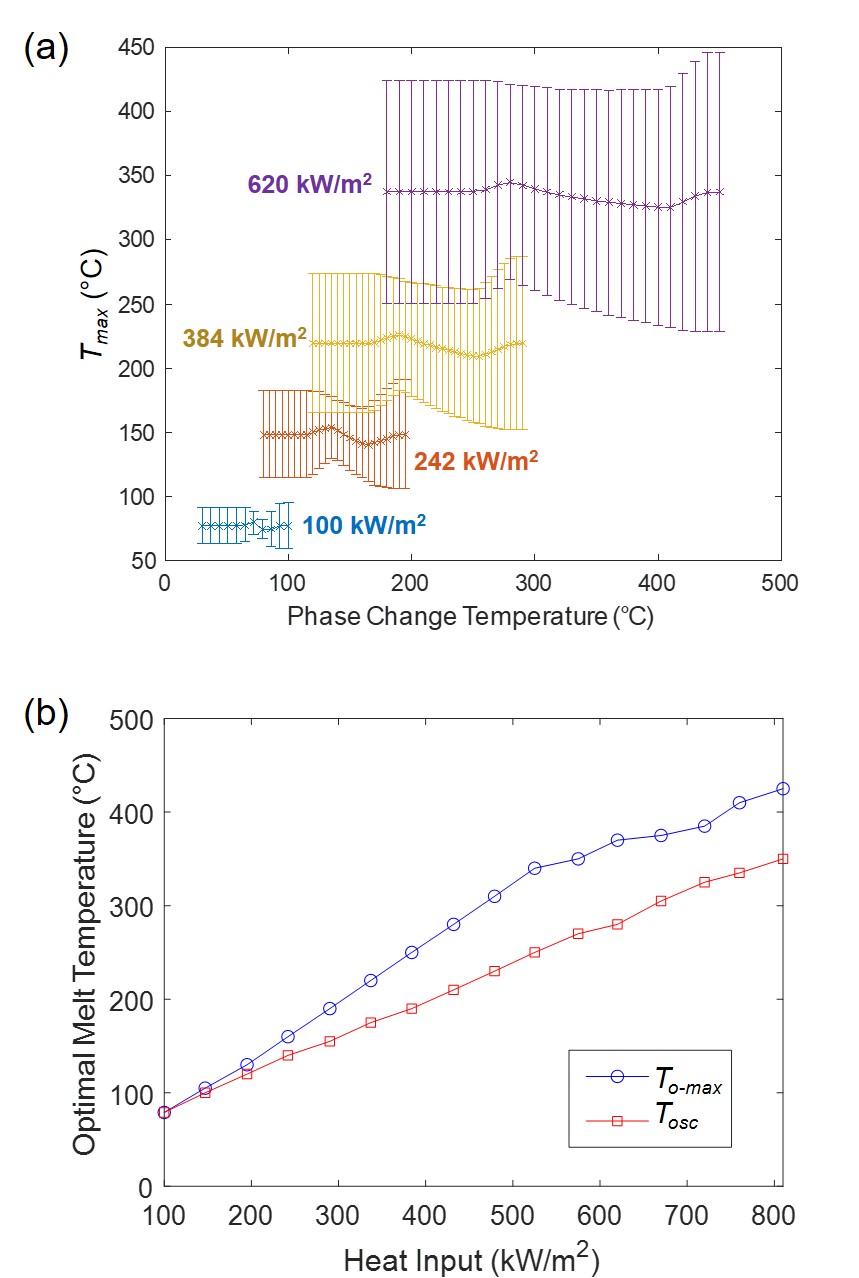}
  \caption{(a) Range of maximum chip temperature during a cycle of heating as a function of PCM melt temperature for four input power levels. Numbers beside the contours indicate their power levels. Apart from $T_{m}$, the properties of Solder 174 are used. As the power levels increase, the separation between the optimal $T_m$ for minimizing the overall maximum chip temperature ($T_{o-max}$, \textit{i.e.}, the top of each bar) deviates from the optimal $T_m$ to minimize the range of oscillation ($T_{osc}$, \textit{i.e.}, the height of each bar). (b) Optimal melt temperature to minimize $T_{o-max}$ and $T_{osc}$ for a range of power inputs.}
  \label{fig:Tmelt}
\end{figure} 

Based on the results of the study of commercial PCMs, the melt temperature appears to strongly impact the cooling performance at a given power level. Building on this, we evaluate the impact of melt temperature on cooling performance for varying input power levels. Apart from the melt temperature, all other thermophysical properties of the PCM match that of the best performing PCM from the first study: Solder 174. Manual sweeping through potential phase change temperatures (with coarse steps) is explored for four different power levels. The range of temperature oscillations for each PCM melt temperature and power level are shown in Fig.~\ref{fig:Tmelt}. For low melt temperatures (the left portion of each curves), the temperature response is independent of $T_{m}$ because the PCM fully melts by the quasi-steady portion of the analysis. It remains fully melted even in the `off' portion of the heating cycle, so the thermal response is purely from conduction and simply the sensible heating response. A plateau is also visible for high PCM melt temperatures (\textit{i.e.}, the right portion of the curves). In this region, $T_m$ is higher than the maximum temperatures in the system and the PCM never melts. Again, there is only conduction and sensible heating response. The dip for intermediate $T_m$ results from melting and solidification in each cycle of heating.

From these parametric sweeps, the optimal melt temperature for minimizing the overall maximum chip temperature and for minimizing the amplitude of the temperature oscillations can be identified. In the case of 100~\si{\kilo \watt \per{\square \meter}}, both metrics are minimized for approximately the same melt temperature ($T_m = $ 77~\si{\celsius}). However, for higher power levels, the optimal melt temperature to minimize the overall maximum temperature is slightly higher than the optimal melt temperature to minimize the range of the temperature oscillations. 

For finer resolution of the optimal $T_m$ values, for a cyclic heat input of 100~\si{\kilo \watt \per{\square \meter}}, we next sweep $T_m$ with 1~\si{\celsius} steps between $T_m$ of 47~\si{\celsius} and 96~\si{\celsius}. Both $T_{o-max}$ and $T_{osc}$ yield an optimum melt temperature of 77~\si{\celsius}. For a single parameter, this approach is relatively quick as the range of possible values within the set boundaries (50 melt temperatures) is small and the whole parameter space can be evaluated within 6 minutes of computational time on a standard desktop computer.

\begin{table*}[t]
\centering
\caption{Optimized $T_m$ and computational time for each optimization strategy. Note that the direct optimization with ParaPower (GA+PP and PSO+PP) require separate computations to optimize for each metric and thus two values are listed, while the NN-assisted optimization requires generation of a training data set which takes the majority of the computational time.  The subscripts and superscripts on the NN-assisted methods indicate the range of optimized parameters based on randomly selecting different training data sets to generate the neural network.}
\begin{tabular}{|c|c|c|c|c|c|}
\hline
    &  & \multicolumn{2}{c|}{\textbf{Optimizing $T_{o-max}$}} & \multicolumn{2}{c|}{\textbf{Optimizing $T_{osc}$}}\\
    \cline{3-6}
    & \textbf{Computational} & \textbf{Optimized} $\mathbf{T_{m}}$ & \textbf{Predicted} $\mathbf{T_{o-max}}$  &  \textbf{Optimized} $\mathbf{T_{m}}$ & \textbf{Predicted} $\mathbf{T_{osc}}$\\
    \textbf{Strategy} & \textbf{Time} & (\si{\celsius}) & (\si{\celsius}) & (\si{\celsius}) & (\si{\celsius}) \\
    \hline
    PS       & 6 min & 77 & 79.43 & 77 & 5.02 \\ \hline
    GA+PP  & \begin{tabular}[c]{@{}c@{}}27 min ($T_{o-max}$)\\ 24 min ($T_{osc}$)\end{tabular} & 77 & 79.43 & 77 & 5.02\\ \hline
    PSO+PP & \begin{tabular}[c]{@{}c@{}}18 min ($T_{o-max}$)\\ 27 min ($T_{osc}$)\end{tabular} & 77 & 79.43 & 77 & 5.02 \\ \hline
    GA+NN  & 6.5 hours & $78_{77}^{79}$  & $80.38_{79.42}^{81.34}$ & $77_{76}^{78}$ & $6.23_{5.02}^{10.76}$\\ \hline
    PSO+NN & 6.5 hours & $78_{77}^{79}$  & $80.38_{79.42}^{81.34}$ & $77_{76}^{77}$ & $6.14_{5.02}^{9.62}$\\ \hline
\end{tabular}
\label{tab:TmOpt}
\end{table*}

\begin{table*}[btp]
\centering
\caption{Optimized Thermophysical Properties and Computational Time for Different Optimization Strategies for ${T_{o-max}}$. The subscripts and superscripts on the NN-assisted methods indicate the range of optimized parameters based on randomly selecting different training data sets to generate the neural network.}
\begin{tabular}{|c|c|c|c|c|c|c|c|}
\hline
\textbf{Strategy} &
\textbf{\begin{tabular}[c]{@{}c@{}}Computational\\ Time\end{tabular}} &
\textbf{\begin{tabular}[c]{@{}c@{}}$\mathbf{T_m}$ \\(\si{\celsius})\end{tabular}}& 
\textbf{\begin{tabular}[c]{@{}c@{}}$\mathbf{L_H}$ \\(\si{\joule \per{\kilo\gram}})\end{tabular}} & 
\textbf{\begin{tabular}[c]{@{}c@{}}$\mathbf{k}$ \\(\si{\watt \per{\meter \kelvin}})\end{tabular}} & 
\textbf{\begin{tabular}[c]{@{}c@{}}$\mathbf{C_{p,solid}}$ \\(\si{\joule \per{\kilo\gram \kelvin}})\end{tabular}} & \textbf{\begin{tabular}[c]{@{}c@{}}$\mathbf{C_{p,liquid}}$ \\(\si{\joule \per{\kilo\gram \kelvin}})\end{tabular}} &
\textbf{\begin{tabular}[c]{@{}c@{}}Predicted\\$\mathbf{T_{o-max}}$ (\si{\celsius}) \end{tabular}}  \\ 
\hline
GA+PP & 4 hours & 78 & 36643 & 35 & 397 & 836 & 80.84 \\ 
\hline
PSO+PP & \begin{tabular}[c]{@{}c@{}} 4.5 hours \end{tabular} & 77 & 47994 & 36 & 401 & 882 & 79.43 \\ 
\hline
GA+NN & 6.5 hours & $78_{78}^{79}$  & $44476_{36184}^{47873}$  & $18_{13}^{34}$  & $382_{350}^{401}$  & $827_{761}^{880}$  & $80.69_{80.38}^{81.44}$ \\ 
\hline
PSO+NN & 6.5 hours & $78_{77}^{79}$  & $47999_{47988}^{48000}$  & $16_{13}^{36}$  & $401_{401}^{401}$  & $881_{862}^{883}$  & $80.31_{79.44}^{81.37}$ \\ 
\hline
\end{tabular}
\label{tab:OptPCMPropTmax}
\end{table*}

\begin{table*}[t!]
\centering
\caption{Optimized Thermophysical Properties and Computational Time for Different Optimization Strategies for ${T_{osc}}$. The subscripts and superscripts with the results for the NN strategies indicate the range of optimized parameters based on randomly selecting different training data sets to generate the neural network. }
\begin{tabular}{|c|c|c|c|c|c|c|c|}
\hline
\textbf{Strategy} &
\textbf{\begin{tabular}[c]{@{}c@{}}Computational\\ Time\end{tabular}} &
\textbf{\begin{tabular}[c]{@{}c@{}}$\mathbf{T_m}$ \\(\si{\celsius})\end{tabular}}& 
\textbf{\begin{tabular}[c]{@{}c@{}}$\mathbf{L_H}$ \\(\si{\joule \per{\kilo\gram}})\end{tabular}} & 
\textbf{\begin{tabular}[c]{@{}c@{}}$\mathbf{k}$ \\(\si{\watt \per{\meter \kelvin}})\end{tabular}} & 
\textbf{\begin{tabular}[c]{@{}c@{}}$\mathbf{C_{p,solid}}$ \\(\si{\joule \per{\kilo\gram \kelvin}})\end{tabular}} & \textbf{\begin{tabular}[c]{@{}c@{}}$\mathbf{C_{p,liquid}}$ \\(\si{\joule \per{\kilo\gram \kelvin}})\end{tabular}} &
\textbf{\begin{tabular}[c]{@{}c@{}}Predicted\\$\mathbf{T_{osc}}$ (\si{\celsius}) \end{tabular}}  \\ 
\hline
GA+PP & 4 hours & 77 & 45190 & 36 & 392 & 883 & 6.18 \\ 
\hline
PSO+PP & \begin{tabular}[c]{@{}c@{}} 6.5 hours \end{tabular} & 77 & 47998 & 36 & 401 & 883 & 5.00 \\ 
\hline
GA+NN & 6.5 hours & $77_{76}^{77}$  & $46609_{44675}^{47713}$  & $17_{13}^{28}$  & $380_{295}^{401}$  & $785_{656}^{873}$  & $6.10_{5.11}^{9.75}$  \\ 
\hline
PSO+NN & 6.5 hours & $77_{77}^{77}$  & $47967_{47669}^{48000}$  & $13_{13}^{18}$  & $401_{401}^{401}$  & $877_{822}^{833}$  & $5.07_{5.03}^{5.10}$ \\ 
\hline
\end{tabular}
\label{tab:OptPCMPropTosc}
\end{table*}

To compare with machine learning assisted optimization strategies, the same cases are evaluated with the direct optimization and the NN-assisted optimization (see Table~\ref{tab:TmOpt}). Using the GA and PSO optimization tools coupled directly to ParaPower, the same optimal $T_m$ = 77~\si{\celsius} is identified, although the computational time is increased to 18 to 27 minutes. This approach takes longer since the algorithm does not terminate immediately even after evaluating 77 ~\si{\celsius}, but continues computing until all convergence criteria are met. Potentially, changing the convergence criteria could reduce the computational time.

 The NN-assisted strategy computes an average optimum melt temperature of 78~\si{\celsius} when optimizing for minimum $T_{o-max}$ and 77~\si{\celsius} when optimizing for minimum $T_{osc}$. The range of $T_{osc}$ predictions show that the sensitivity of $T_m$ on system performance is very high even for a 1~\si{\celsius} change. This strategy takes 6.5 hours, which includes 6.5 hours to generate the training data, 10 seconds to train the neural network, and $<$1 second to compute the optimal value. The time limiting step is generating the training data set. Reducing the training data set size could improve the efficiency of this algorithm, but small training sets may not lead to accurate NN models for optimization. 
 
In the case of optimizing a single parameter, the parametric sweep is sufficient and recommended given the limited search space. But manual optimization is limited to sweeping a few parameters or evaluating a limited number of data points. This motivates the study to explore other strategies including those that can efficiently optimize multiple parameters simultaneously.

\subsection{Optimizing the Thermophysical Properties for a Fixed Geometry}

In this phase of the study, we consider optimization of five properties of the PCM: 

\begin{enumerate}
\item $T_{m}$ (range: 47 - 96 ~\si{\celsius})
\item $L_H$ (range: 25000 - 48000 J/kg)
\item $k$ (range: 10 - 36~\si{\joule \per{\kilo\gram}})
\item $c_{p,solid}$ (range: 146 - 401~\si{\joule \per{\kilo\gram \kelvin}})
\item $c_{p,liquid}$ (range: 167 - 883~\si{\joule \per{\kilo\gram \kelvin}})
\end{enumerate}
The geometry is fixed to have a channel with $H = $ 100 ~\si{\micro\meter} and $W = $ 50 ~\si{\micro\meter}. The parametric search was deemed infeasible for this study as an exhaustive search would take approximately 445,700 years to compute with the desired resolution in each parameter. The results of the optimization study are shown in table \ref{tab:OptPCMPropTmax} and \ref{tab:OptPCMPropTosc}. 

The results show that there exists a unique optimum PCM melt temperature for different systems, and that remaining PCM properties must simply be maximized for best performance. The optimal melt temperature ranges from 76~\si{\celsius} to 79~\si{\celsius} similar to the results of the single parameter optimization. In general, $L_H$, $c_{p,solid}$, and $c_{p,liquid}$ approximately saturate at the maximum value of the search range for most of the optimization strategies. This is expected intuitively as higher energy storage capacity (maximizing $L_H$, $c_{p,solid}$, and $c_{p,liquid}$) improves heat sinking capability. 

In one case, the GA+PP optimization, shows an optimal latent heat less than the maximum end of the allowed range. It is possible that when the GA mutated or crossed over the latent heat towards the upper bound in a specific child, the remaining properties within that set could have caused its performance to be relatively weaker. Thus, this child would not have been allowed to crossover to the next generation. 

\begin{figure}[tb!]
  \includegraphics[width=\linewidth]{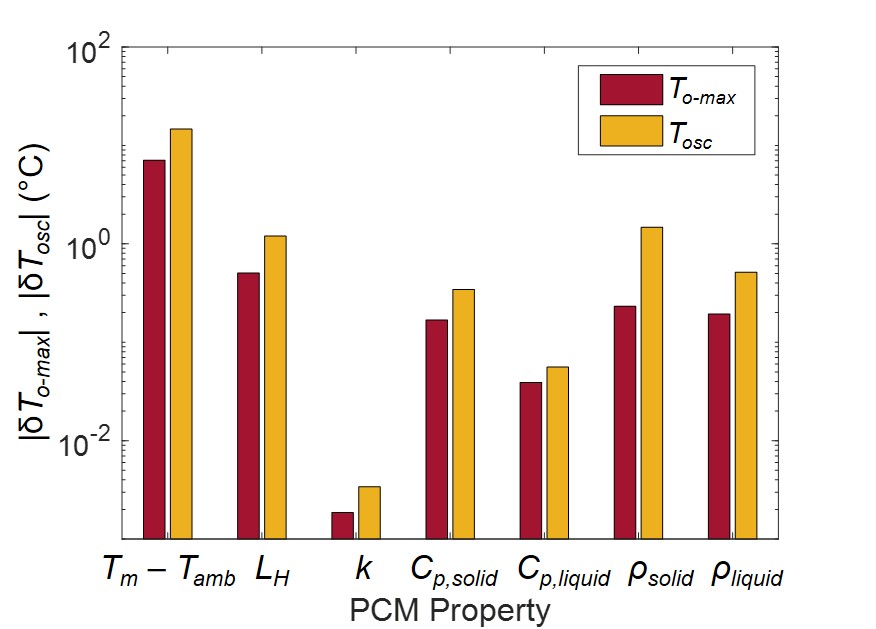}
  \caption{Sensitivity of $T_{o-max}$ ($|\delta T_{o-max}|$, shown in red) and $T_{osc}$ ($|\delta T_{osc}|$, shown in orange) to different PCM properties. Each property is increased and decreased by 10\% while holding the remaining properties constant, and the average change in both metrics is reported here. The impact of $T_m$ is adjusted by changing $T_m - T_{amb}$ by 10\% rather than 10\% of $T_m$ directly. The melt temperature has the greatest impact on the system performance.}
  \label{fig:Sensitivity}
\end{figure} 
\begin{table*}[b!]
\centering

\caption{Optimized Channel Height, Channel Width, and Melt Temperature, as well as Required Computational Time, for Different Optimization Strategies. Computational time for the NN assisted optimization depends on the training data set size: 2380 data sets are used for the 4.5 hours of computational time and 250 data sets are used for the 0.47 hours of computational time. The subscripts and superscripts indicate the range of optimized parameters based on randomly selecting different training data sets to generate the neural network. }
\begin{tabular}{|c|c|c|c|c|c|c|c|c|c|}
\hline
{\textbf{Strategy} } & 
{\begin{tabular}[c]{@{}c@{}}\textbf{Computational}\\\textbf{ Time} \end{tabular}} & 
  \multicolumn{3}{c|}{\textbf{Optimizing} $\mathbf{T_{o-max}}$} & {\begin{tabular}[c]{@{}c@{}}\textbf{ Predicted}\\\textbf{$\mathbf{T_{o-max}}$ } \end{tabular}} & 
  \multicolumn{3}{c|}{\textbf{Optimizing} $\mathbf{T_{osc}}$} & {\begin{tabular}[c]{@{}c@{}}\textbf{Predicted }\\\textbf{$\mathbf{T_{osc}}$ } \end{tabular}}\\ 
\cline{3-5}\cline{7-9}
 &  &
 \begin{tabular}[c]{@{}c@{}}$H$\\ ({$\mu$}m) \end{tabular} & 
 \begin{tabular}[c]{@{}c@{}}$W$ \\ ({$\mu$}m) \end{tabular} & 
 \begin{tabular}[c]{@{}c@{}}$T_m$ \\ (~\si{\celsius}) \end{tabular} &  (~\si{\celsius})
 & 
 \begin{tabular}[c]{@{}c@{}}$H$\\ ({$\mu$}m) \end{tabular} & 
 \begin{tabular}[c]{@{}c@{}}$W$ \\ ({$\mu$}m) \end{tabular} & 
 \begin{tabular}[c]{@{}c@{}}$T_m$ \\ (~\si{\celsius}) \end{tabular} &  (~\si{\celsius})\\ 
\hline
    GA+PP   & \begin{tabular}[c]{@{}l@{}}2.5 hours ($T_{o-max}$)\\ 2.9 hours ($T_{osc}$)\end{tabular} & 100     & 99    & 77 & 78.5  & 100    & 100    & 77 & 2.9   \\ \hline
    PSO+PP  & \begin{tabular}[c]{@{}l@{}}2.3 hours ($T_{o-max}$)\\ 4.5 hours ($T_{osc}$)\end{tabular} & 100     & 97    & 77 & 78.5    & 100    & 100    & 77 & 2.9   \\ \hline
    GA+NN  & 4.5 hours & $95_{83}^{100}$  & $76_{60}^{100}$  & $78_{78}^{78}$  & $79.9_{79.4}^{80.1}$  & $96_{75}^{100}$  & $91_{67}^{100}$  & $77_{76}^{77}$  & $4.1_{2.9}^{8.6}$ \\  
    \hline
    PSO+NN  & 4.5 hours & $95_{75}^{100}$  & $79_{62}^{100}$  & $78_{78}^{78}$ & $79.8_{79.4}^{80.1}$    & $100_{100}^{100}$  & $90_{66}^{100}$  & $77_{76}^{77}$ & $4.0_{2.9}^{8.6}$  \\
    \hline
    GA+NN   & 0.47 hours   & $91_{67}^{100}$  & $89_{64}^{100}$  & $79_{78}^{79}$  & $80.2_{79.4}^{81.4}$  & $92_{68}^{100}$  & $85_{64}^{100}$  & $77_{76}^{78}$  & $6.1_{2.9}^{8.0}$ \\  
    \hline
    PSO+NN  & 0.47 hours & $85_{51}^{100}$  & $98_{83}^{100}$  & $79_{78}^{80}$ & $80.3_{79.4}^{81.4}$    & $93_{66}^{100}$  & $92_{72}^{100}$  & $77_{76}^{78}$ & $5.9_{2.9}^{9.1}$  \\
\hline
\end{tabular}
\label{tab:GeomTmOpt}
\end{table*}

In the NN-assisted strategies, the thermal conductivity $k$ averages near the minimum value of the search range. Although better thermal transport (higher $k$) should improve heat sinking capability, its impact on the performance is relatively lower than that of the other PCM properties. The sensitivity of $T_{o-max}$ or $T_{osc}$ (defined as $\delta T_{o-max}$ and $\delta T_{osc}$ respectively) with regards to the the five PCM properties was explored by adjusting each property by 10\% and observing the change in the performance parameters (Fig.~\ref{fig:Sensitivity}). The impact of $k$ on the two metrics is several orders of magnitude lower than the other PCM properties. As a result, the NN fails to accurately capture the effect of thermal conductivity. Even though the average predicted $k$ optimum for the NN-assisted optimization is less than half of that predicted by the ParaPower assisted optimization, the average performance of the PCMs (consisting of all the 5 properties) for both strategies is similar.

Overall, with the multi-parameter optimization, coupling the optimization algorithms directly to ParaPower (GA+PP and PSO+PP) are more computationally efficient than the neural network approach because an optimum can be found in less iterations that training the neural network. The direct optimization strategies take about 5 hours compared to the 6.5 hours required to train the data set for the neural network. Again, the NN-assisted optimization could be sped up at a cost of accuracy by reducing the size of the NN training data set (see section 3.5). Here, intuition could be used to reduce the optimization to focusing just on melt temperature rather than 5 independent thermophysical properties, but this study begins to illustrate the relative advantages and disadvantages of the different optimization strategies.

\subsection{Optimizing the Channel Geometry ($H$ and $W$) and Melt Temperature}

Finally, we consider simultaneous optimization of the melt temperature and the height and width of the PCM channel. The remaining PCM properties are fixed at that of solder 174 since that was the best performing PCM in the initial study. The parametric search is again infeasible as an exhaustive search of the parameter space is estimated to take $\sim$28 days. Results from the machine learning-assisted optimization strategies are shown in Table~\ref{tab:GeomTmOpt}.

When optimizing to minimize the range of temperature oscillations $T_{osc}$, ParaPower assisted strategies predict that the PCM channel should fill the entire allowed zone ($H = $ 100~\si{\micro\meter} and $W = $ 100~\si{\micro\meter}) with a PCM melt temperature of 77~\si{\celsius}. Similarly, when minimizing overall maximum chip temperature $T_{o-max}$, both PP-based strategies predict that the PCM channel should fill nearly the entire allowed zone ($H = $ 100~\si{\micro\meter} and $W \sim $ 97 to 99~\si{\micro\meter}) with a PCM melt temperature of 77~\si{\celsius}. This leaves a thin channel for conduction to the heat sink between the PCM channels. The channels maximizing the allowed volume may be in part due to the uniform heat source and the relatively high thermal conductivity of the PCM. More complex cases and localized heating may yield different optimum geometries.

NN-assisted strategies similarly predict the PCM pocket should nearly fill the layer to minimize $T_{osc}$ and $T_{o-max}$ with PCM melt temperatures ranging between 76~\si{\celsius} and 78~\si{\celsius}. Interestingly, the average optimized channel width for minimizing $T_{o-max}$ is much lower than the almost-filled width predicted by the ParaPower assisted strategy. But when minimizing $T_{osc}$, the average optimized width is closer to the full channel width. In part, these variations are due to the generation of the neural network: with the NN-assisted approaches, the optimization results depend on the data sets selected for training the neural network. The subscripts and superscripts next to the values indicate the range of optimized parameters repeating the NN generation and optimization for 10 randomly selected data sets. The results also depends on the number of training data points used to generate the neural networks. Table~\ref{tab:GeomTmOpt} shows results for 2380 data sets (4.5 hours of computational time) and 250 data sets (0.47 hours of computational time). Once the NN is generated optimization for $T_{o-max}$ and $T_{osc}$ is minimal compared to the time required to generate the training data in ParaPower. This trend is further explored in the following section. 

Overall, the direct optimization with the algorithms coupled to ParaPower is again the most effective solution method. However, using this strategy for different sets of operating conditions or target metrics requires re-starting the optimization. Thus, times are given for the PP-optimizations with minimizing $T_{o-max}$ and for minimizing $T_{osc}$. On the other hand, with the NN-assisted optimization, once the training data is generated, it can be used to optimize for different metrics on demand.

\begin{table*}[bt!] 
\centering
\caption{Trends in NN accuracy and ML predicted optimums for $T_{o-max}$. The subscripts and superscripts indicate the range of optimized parameters based on randomly selecting different training data sets to generate the neural network.}
\label{tab:NNTrendsMax}
\begin{tabular}{|c|c|c|c|c|c|c|c|c|c|c|} 
    \hline
    {\begin{tabular}[c]{@{}c@{}}\textbf{Computational}\\ \textbf{Time} \end{tabular}} & 
    {\begin{tabular}[c]{@{}c@{}}\textbf{Training}\\ \textbf{Data}\\ \textbf{Size}\end{tabular}} & 
    {$\mathbf{R^{2}}$} &
    \multicolumn{3}{c|}{\textbf{PSO}} & 
    {\begin{tabular}[c]{@{}c@{}}\textbf{ Predicted}\\\textbf{$\mathbf{T_{o-max}}$ } \end{tabular}} &
    \multicolumn{3}{c|}{\textbf{GA}} &
    {\begin{tabular}[c]{@{}c@{}}\textbf{ Predicted}\\\textbf{$\mathbf{T_{o-max}}$ } \end{tabular}} \\ 
    \cline{4-6}\cline{8-10}
    
    (hrs) &  &  & 
     \begin{tabular}[c]{@{}c@{}}$H$\\ ({$\mu$}m) \end{tabular} & 
     \begin{tabular}[c]{@{}c@{}}$W$ \\ ({$\mu$}m) \end{tabular} & 
     \begin{tabular}[c]{@{}c@{}}$T_m$ \\ (~\si{\celsius}) \end{tabular} &  (~\si{\celsius})
     & 
     \begin{tabular}[c]{@{}c@{}}$H$\\ ({$\mu$}m) \end{tabular} & 
     \begin{tabular}[c]{@{}c@{}}$W$ \\ ({$\mu$}m) \end{tabular} & 
     \begin{tabular}[c]{@{}c@{}}$T_m$ \\ (~\si{\celsius}) \end{tabular} &  (~\si{\celsius})\\ 
    \hline

    0.06 & 30 & $0.37_{0.00}^{0.76}$  & $94_{35}^{100}$  & $75_{58}^{100}$  & $76_{47}^{96}$ & $86.4_{78.8}^{94.0}$ & $84_{48}^{100}$  & $69_{20}^{100}$  & $77_{50}^{95}$  & $87.0_{82.2}^{93.9}$  \\ 
    \hline
    0.08 & 40 & $0.50_{0.14}^{0.86}$  & $87_{20}^{100}$  & $60_{20}^{100}$  & $83_{70}^{96}$ & $87.5_{79.6}^{96.2}$ & $93_{68}^{100}$  & $74_{20}^{100}$  & $82_{76}^{93}$  & $84.5_{78.9}^{92.8}$  \\ 
    \hline
    0.09 & 50 & $0.62_{0.02}^{0.89}$  & $90_{44}^{100}$  & $81_{26}^{100}$  & $75_{47}^{82}$ & $84.1_{79.7}^{93.2}$ & $89_{66}^{100}$  & $80_{42}^{100}$  & $73_{47}^{82}$  & $83.4_{79.1}^{93.1}$  \\ 
    \hline
    0.19 & 100 & $0.90_{0.71}^{0.95}$  & $94_{56}^{100}$  & $77_{51}^{100}$  & $79_{77}^{82}$ & $80.6_{78.8}^{83.3}$ & $88_{60}^{100}$  & $83_{65}^{100}$  & $79_{77}^{83}$  & $80.7_{79.0}^{84.3}$  \\ 
    \hline
    0.28 & 150 & $0.96_{0.93}^{0.98}$  & $87_{52}^{100}$  & $88_{66}^{100}$  & $78_{75}^{79}$ & $80.6_{78.4}^{84.3}$ & $78_{54}^{100}$  & $89_{68}^{100}$  & $79_{78}^{80}$  & $80.5_{79.4}^{82.0}$  \\ 
    \hline
    0.47 & 250 & $0.97_{0.95}^{0.99}$  & $85_{51}^{100}$  & $98_{83}^{100}$ & $79_{78}^{80}$ &  $80.3_{79.4}^{81.4}$ & $91_{67}^{100}$  & $89_{64}^{100}$  & $79_{78}^{80}$  & $80.2_{79.4}^{81.4}$  \\ 
    \hline
    0.95 & 500 & $0.98_{0.97}^{0.99}$  & $96_{66}^{100}$  & $87_{65}^{100}$  & $78_{77}^{80}$ & $80.1_{78.4}^{81.4}$  & $84_{62}^{100}$  & $94_{82}^{100}$  & $78_{78}^{79}$  & $80.0_{79.4}^{80.9}$   \\ 
    \hline
    1.89 & 1000 & $0.98_{0.95}^{0.99}$  & $92_{63}^{100}$  & $92_{69}^{100}$  & $78_{78}^{78}$ & $79.6_{79.4}^{80.1}$  & $94_{78}^{100}$  & $88_{70}^{100}$  & $78_{78}^{79}$  & $79.8_{79.4}^{80.8}$   \\ 
    \hline
    3.78 & 2000 & $0.99_{0.99}^{0.99}$  & $97_{75}^{100}$  & $91_{62}^{100}$  & $78_{77}^{79}$ & $79.5_{78.4}^{80.4}$  & $91_{68}^{100}$  & $93_{71}^{100}$  & $78_{77}^{79}$  & $79.6_{78.4}^{80.4}$   \\ 
    \hline
    4.50 & 2380 & $0.99_{0.98}^{0.99}$  & $95_{75}^{100}$  & $79_{62}^{100}$  & $78_{78}^{78}$ & $79.8_{79.4}^{80.1}$  & $95_{83}^{100}$  & $76_{60}^{100}$  & $78_{78}^{78}$  & $79.9_{79.4}^{80.1}$   \\
    \hline
\end{tabular}
\end{table*}

\begin{table*}[bt!] 
\centering
\caption{Trends in NN accuracy and ML predicted optimums for $T_{osc}$. The subscripts and superscripts indicate the range of optimized parameters based on randomly selecting different training data sets to generate the neural network.}
\label{tab:NNTrendsOsc}
\begin{tabular}{|c|c|c|c|c|c|c|c|c|c|c|} 
    \hline
    {\begin{tabular}[c]{@{}c@{}}\textbf{Computational}\\ \textbf{Time} \end{tabular}} &
    {\begin{tabular}[c]{@{}c@{}}\textbf{Training}\\ \textbf{Data}\\ \textbf{Size}\end{tabular}} & 
    {$\mathbf{R^{2}}$} &
    \multicolumn{3}{c|}{\textbf{PSO}} & 
    {\begin{tabular}[c]{@{}c@{}}\textbf{ParaPower}\\\textbf{ Predicted}\\\textbf{$\mathbf{T_{osc}}$} \end{tabular}} &
    \multicolumn{3}{c|}{\textbf{GA}} &
    {\begin{tabular}[c]{@{}c@{}}\textbf{ParaPower}\\\textbf{ Predicted}\\\textbf{$\mathbf{T_{osc}}$} \end{tabular}} \\ 
    \cline{4-6}\cline{8-10}
    
     (hrs) &  &  & 
     \begin{tabular}[c]{@{}c@{}}$H$\\ ({$\mu$}m) \end{tabular} & 
     \begin{tabular}[c]{@{}c@{}}$W$ \\ ({$\mu$}m) \end{tabular} & 
     \begin{tabular}[c]{@{}c@{}}$T_m$ \\ (~\si{\celsius}) \end{tabular} &  (~\si{\celsius})
     & 
     \begin{tabular}[c]{@{}c@{}}$H$\\ ({$\mu$}m) \end{tabular} & 
     \begin{tabular}[c]{@{}c@{}}$W$ \\ ({$\mu$}m) \end{tabular} & 
     \begin{tabular}[c]{@{}c@{}}$T_m$ \\ (~\si{\celsius}) \end{tabular} &  (~\si{\celsius})\\ 
    \hline
    
    0.06 & 30 & $0.52_{0.24}^{0.78}$  & $89_{20}^{100}$  & $81_{20}^{100}$  & $78_{52}^{96}$ & $18.3_{3.4}^{34.5}$ & $97_{74}^{100}$  & $84_{62}^{100}$  & $81_{70}^{96}$  & $17.1_{3.5}^{33.0}$  \\ 
    \hline
    0.08 & 40 & $0.39_{0.12}^{0.69}$  & $83_{20}^{100}$  & $81_{20}^{100}$  & $84_{64}^{96}$ & $26.6_{12.7}^{36.9}$ & $92_{74}^{100}$  & $83_{58}^{100}$  & $82_{72}^{96}$  & $22.7_{9.0}^{36.9}$  \\ 
    \hline
    0.09 & 50 & $0.58_{0.08}^{0.83}$  & $95_{74}^{100}$  & $80_{20}^{100}$  & $80_{47}^{96}$ & $19.4_{3.9}^{34.4}$ & $92_{74}^{100}$  & $83_{58}^{100}$  & $82_{72}^{96}$  & $17.8_{3.5}^{32.9}$  \\ 
    \hline
    0.19 & 100 & $0.89_{0.70}^{0.95}$  & $96_{65}^{100}$  & $96_{70}^{100}$  & $76_{74}^{78}$ & $8.1_{2.9}^{12.1}$ & $87_{64}^{100}$  & $93_{69}^{100}$  & $77_{74}^{78}$  & $7.5_{2.9}^{12.5}$  \\ 
    \hline
    0.28 & 150 & $0.95_{0.87}^{0.97}$  & $94_{58}^{100}$  & $89_{67}^{100}$  & $77_{76}^{77}$ & $5.7_{2.9}^{9.0}$ & $91_{55}^{100}$  & $89_{72}^{100}$  & $77_{76}^{77}$  & $5.3_{2.9}^{9.2}$  \\ 
    \hline
    0.47 & 250 & $0.96_{0.93}^{0.97}$  & $93_{66}^{100}$  & $92_{72}^{100}$ & $77_{76}^{78}$ &  $5.9_{2.9}^{9.1}$ & $92_{68}^{100}$  & $85_{64}^{100}$  & $77_{76}^{78}$  & $6.1_{3.0}^{10.0}$  \\ 
    \hline
    0.95 & 500 & $0.98_{0.96}^{0.99}$  & $92_{73}^{100}$  & $93_{68}^{100}$  & $77_{76}^{77}$ & $5.6_{2.9}^{8.5}$  & $94_{73}^{100}$  & $90_{61}^{100}$  & $77_{76}^{77}$  & $4.2_{2.9}^{8.0}$   \\ 
    \hline
    1.89 & 1000 & $0.99_{0.98}^{0.99}$  & $97_{67}^{100}$  & $92_{66}^{100}$  & $77_{76}^{78}$ & $7.6_{2.9}^{9.1}$  & $94_{88}^{100}$  & $89_{70}^{100}$  & $77_{76}^{78}$  & $5.8_{2.9}^{9.1}$   \\ 
    \hline
    3.78 & 2000 & $0.99_{0.99}^{0.99}$  & $95_{66}^{100}$  & $82_{69}^{100}$  & $77_{76}^{77}$ & $5.8_{2.9}^{8.6}$  & $92_{79}^{100}$  & $83_{64}^{100}$  & $77_{76}^{77}$  & $5.9_{2.9}^{8.8}$   \\ 
    \hline
    4.50 & 2380 & $0.99_{0.97}^{0.99}$  & $100_{100}^{100}$  & $90_{66}^{100}$  & $77_{76}^{77}$ & $4.0_{2.9}^{8.6}$  & $96_{75}^{100}$  & $91_{67}^{100}$  & $77_{76}^{77}$  & $4.1_{2.9}^{8.6}$   \\
    \hline
\end{tabular}
\end{table*}

\subsection{Impact of Training Data Set Size on NN-assisted Optimization}

The accuracy of the results in NN-assisted optimization strategies are heavily influenced by the accuracy of the NN when compared to the optimization algorithm and its selected convergence criteria. The performance of a NN can be impacted by the structure of a NN (number of hidden layers, neurons per layer, activation function), amount of training data available and the randomly initialized weights. The impact of the training data set size was explored on the previous optimization objective (optimizing geometry and PCM melt temperature). 

\begin{figure}[th!]
  \includegraphics[width=\linewidth]{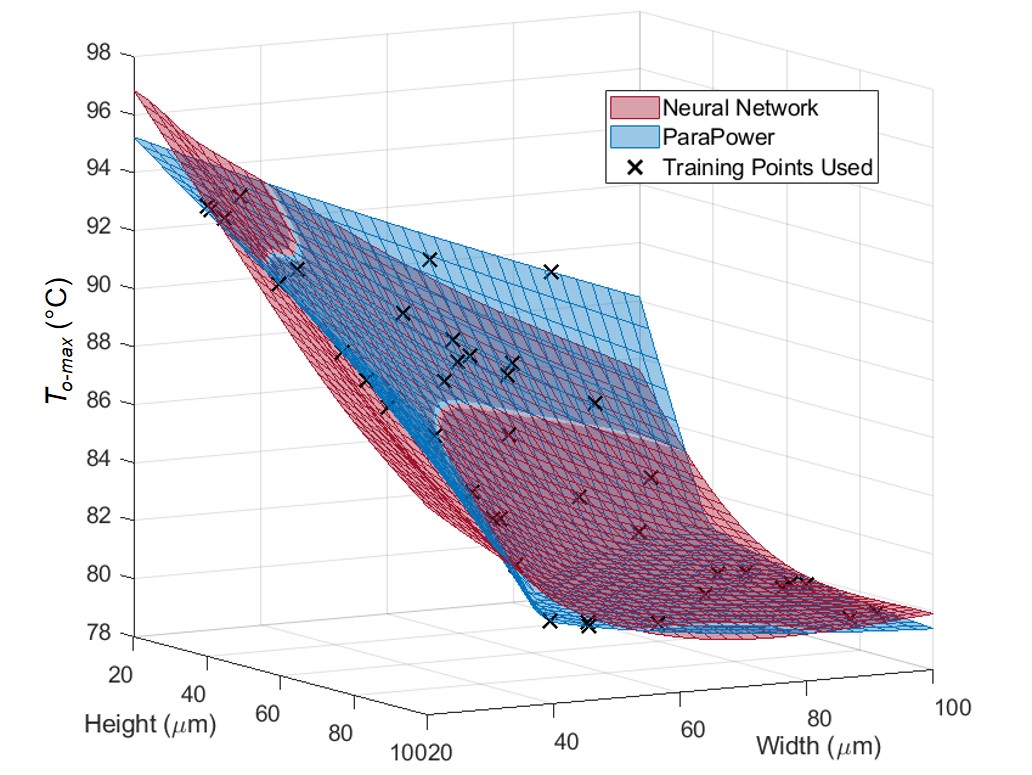}
  \caption{$T_{o-max}$ predictions by a NN (shown in red) alongside ParaPower predictions (shown in blue) for a range of width and height values, and a fixed $T_m$ of 78~\si{\celsius}. The black crosses represent values used for training the NN from within the set of 2380 training cases. Unlike the ParaPower predictions, the NN has a distinct minima at a height of 100~\si{\micro\meter} and width of 75~\si{\micro\meter}. These difference likely account for differences in predicted optima with the NN versus directly through ParaPower.}
  \label{fig:NNWidthVariation}
\end{figure} 

The available training data set of 3300 cases was divided into 10 different sizes. Then, the NN was trained and GA+NN and PSO+NN strategies were run for each training data size 10 times. This way, both the impact of the training data size, and the randomly initialized weights was observed. To test the accuracy of a NN, the $R^{2}$ value of its fit was calculated through a separate test set of 500 cases. The results of this study are compiled in tables ~\ref{tab:NNTrendsMax} and ~\ref{tab:NNTrendsOsc}. The average values of the $R^{2}$ and predicted optimums for each NN are shown. The superscript and subscript next to each metric denote the maximum and minimum value observed for the 10 runs.

Increasing the training data is seen to improve the performance of the NN. However, this benefit begins to slowly saturate past a training data size of 250 cases and the difference in NN accuracies between 2000 training data values and 2380 training data values is minimal. Increasing the amount of training improves the accuracy of the NN-based approaches at the penalty of computatioanl time. An interesting observation is seen in the predicted width optimum (for both $T_{o-max}$ and $T_{osc}$) – the fluctuations in the optimum predictions are still large despite relatively high $R^{2}$ values. This could be due to overfitting of the NN in the range of values that are not near the optimum zones. 

This hypothesis was tested by checking the spread of $T_{o-max}$ over varying channel heights and widths (for a fixed $T_{m}$ of 78~\si{\celsius} ). Figure ~\ref{fig:NNWidthVariation} shows $T_{o-max}$ predictions by a NN (red) alongside those by ParaPower (blue). The NN predictions do match the general shape of the ParaPower predictions, but unlike the ParaPower surface with a continued downward slope, the NN predicted surface has a distinct minima at a height of 100 ~\si{\micro\meter} and width of 75 ~\si{\micro\meter}. This is why the optimization algorithms GA and PSO predict a width optima that is not near the upper bound.  The relative flatness of the ParaPower surface past a height and width of \textasciitilde 60~\si{\micro\meter} show that the system performance is not very sensitive beyond this point. Thus, despite this inaccuracy, the average ParaPower predicted $T_{o-max}$ is still very similar (\textasciitilde  1~\si{\celsius} difference) to those predicted by the GA+PP and PSO+PP strategies from table ~\ref{tab:GeomTmOpt}. Ultimately the plateau gives a large region of design space with good thermal performance.
\section{Conclusion}
In this work, we demonstrated the effectiveness of  thermal management using PCMs embedded within the electronic chip. In a fixed geometry, Solder 174 demonstrated the highest performance (reduced the maximum temperature rise by 19\% and the amplitude of the temperature oscillations by 88\%, with the system never reaching the cutoff of 85~\si{\celsius}) of all the commercially-available PCMs investigated. Initial optimization of just the melt temperature as a function of heater power demonstrated increasing optimal melt temperature with increasing heater power. There is also a tradeoff between minimizing $T_{o-max}$ and $T_{osc}$, with the optimum $T_m$ to minimize $T_{o-max}$ being increasingly higher than the optimum $T_m$ for minimizing $T_{osc}$ with increasing heater power. In the multi-parameter optimization of the channel geometry and PCM properties, (i) the optimized designs generally had small solid channels between large pockets of the solder 174-based PCM channels and (ii) maximizing all thermophyscial properties other than melt temperature ($k$, $C_{p,solid}$, $C_{p,liquid}$, $L_h$) led to optimal performance.

In the multi-parameter optimizations, all machine learning strategies converged on system optimums that would have otherwise taken several days for manual searches. Direct optimization using the GA and PSO algorithms coupled to ParaPower was efficient at achieving the optimization compared to predicted time to compute parametric sweeps. The computational time for NN-assisted optimization depends on the size of the data set used for training. For this system, a minimum of 250 data sets ($\sim$ 0.5 hrs computational time) is required for relatively consistent optimized results. In comparison, the direct optimization routines took several hours to reach an optimum. The neural network once built adds flexibility as it can  be reused for optimization of multiple parameters.

Embedded PCM cooling is a promising thermal management solution for electronic devices such as mobile phones that have space and weight constraints and the machine-learning based optimization strategies provide a useful tool for designing systems leveraging these materials. Future work focused on 3-dimensional architectures can include more complex geometric optimization to ensure a balance of thermal buffering PCM zones alongside highly conductive silicon pathways to the ambient. 

\section{acknowledgment}
Financial support for this work provided by members of the Cooling Technologies Research Center, a graduated National Science Foundation Industry/University Cooperative Research Center at Purdue University, is gratefully acknowledged.


 \bibliographystyle{elsarticle-num} 
 \bibliography{myreferences.bib}

\begin{thebibliography}{10}
\expandafter\ifx\csname url\endcsname\relax
  \def\url#1{\texttt{#1}}\fi
\expandafter\ifx\csname urlprefix\endcsname\relax\def\urlprefix{URL }\fi
\expandafter\ifx\csname href\endcsname\relax
  \def\href#1#2{#2} \def\path#1{#1}\fi

\bibitem{Kandasamy2007ApplicationElectronics}
R.~Kandasamy, X.-Q. Wang, A.~S. Mujumdar,
  \href{https://linkinghub.elsevier.com/retrieve/pii/S1359431107000063}{{Application
  of phase change materials in thermal management of electronics}}, Applied
  Thermal Engineering 27~(17-18) (2007) 2822--2832.
\newblock \href {https://doi.org/10.1016/j.applthermaleng.2006.12.013}
  {\path{doi:10.1016/j.applthermaleng.2006.12.013}}.

\bibitem{Tomizawa2016ExperimentalDevices}
Y.~Tomizawa, K.~Sasaki, A.~Kuroda, R.~Takeda, Y.~Kaito,
  \href{https://linkinghub.elsevier.com/retrieve/pii/S1359431115014374}{{Experimental
  and numerical study on phase change material (PCM) for thermal management of
  mobile devices}}, Applied Thermal Engineering 98 (2016) 320--329.
\newblock \href {https://doi.org/10.1016/j.applthermaleng.2015.12.056}
  {\path{doi:10.1016/j.applthermaleng.2015.12.056}}.

\bibitem{Kamkari2018ExperimentalAngles}
B.~Kamkari, D.~Groulx,
  \href{https://doi.org/10.1016/j.expthermflusci.2018.04.007
  https://linkinghub.elsevier.com/retrieve/pii/S0894177718305417}{{Experimental
  investigation of melting behaviour of phase change material in finned
  rectangular enclosures under different inclination angles}}, Experimental
  Thermal and Fluid Science 97~(December 2017) (2018) 94--108.
\newblock \href {https://doi.org/10.1016/j.expthermflusci.2018.04.007}
  {\path{doi:10.1016/j.expthermflusci.2018.04.007}}.

\bibitem{Arshad2018ExperimentalDiameter}
A.~Arshad, H.~M. Ali, S.~Khushnood, M.~Jabbal,
  \href{https://doi.org/10.1016/j.ijheatmasstransfer.2017.10.008
  https://linkinghub.elsevier.com/retrieve/pii/S0017931017328107}{{Experimental
  investigation of PCM based round pin-fin heat sinks for thermal management of
  electronics: Effect of pin-fin diameter}}, International Journal of Heat and
  Mass Transfer 117 (2018) 861--872.
\newblock \href {https://doi.org/10.1016/j.ijheatmasstransfer.2017.10.008}
  {\path{doi:10.1016/j.ijheatmasstransfer.2017.10.008}}.

\bibitem{Ganatra2018ExperimentalDevices}
Y.~Ganatra, J.~Ruiz, J.~A. Howarter, A.~Marconnet,
  \href{https://linkinghub.elsevier.com/retrieve/pii/S1290072917305574}{{Experimental
  investigation of Phase Change Materials for thermal management of handheld
  devices}}, International Journal of Thermal Sciences 129 (2018) 358--364.
\newblock \href {https://doi.org/10.1016/j.ijthermalsci.2018.03.012}
  {\path{doi:10.1016/j.ijthermalsci.2018.03.012}}.

\bibitem{Ji2018Non-uniformFins}
C.~Ji, Z.~Qin, Z.~Low, S.~Dubey, F.~H. Choo, F.~Duan,
  \href{https://doi.org/10.1016/j.applthermaleng.2017.10.030
  https://linkinghub.elsevier.com/retrieve/pii/S1359431117356806}{{Non-uniform
  heat transfer suppression to enhance PCM melting by angled fins}}, Applied
  Thermal Engineering 129 (2018) 269--279.
\newblock \href {https://doi.org/10.1016/j.applthermaleng.2017.10.030}
  {\path{doi:10.1016/j.applthermaleng.2017.10.030}}.

\bibitem{Kalbasi2019StudiesSinks}
R.~Kalbasi, M.~Afrand, J.~Alsarraf, M.-D. Tran,
  \href{https://doi.org/10.1016/j.energy.2019.01.070
  https://linkinghub.elsevier.com/retrieve/pii/S0360544219300726}{{Studies on
  optimum fins number in PCM-based heat sinks}}, Energy 171 (2019) 1088--1099.
\newblock \href {https://doi.org/10.1016/j.energy.2019.01.070}
  {\path{doi:10.1016/j.energy.2019.01.070}}.

\bibitem{Wang2011Three-dimensionalSink}
Y.-H. Wang, Y.-T. Yang, \href{http://dx.doi.org/10.1016/j.energy.2011.06.023
  https://linkinghub.elsevier.com/retrieve/pii/S0360544211004087}{{Three-dimensional
  transient cooling simulations of a portable electronic device using PCM
  (phase change materials) in multi-fin heat sink}}, Energy 36~(8) (2011)
  5214--5224.
\newblock \href {https://doi.org/10.1016/j.energy.2011.06.023}
  {\path{doi:10.1016/j.energy.2011.06.023}}.

\bibitem{Krishnan2004ThermalSinks}
S.~Krishnan, S.~V. Garimella,
  \href{https://asmedigitalcollection.asme.org/electronicpackaging/article/126/3/308/444380/Thermal-Management-of-Transient-Power-Spikes-in}{{Thermal
  Management of Transient Power Spikes in Electronics—Phase Change Energy
  Storage or Copper Heat Sinks?}}, Journal of Electronic Packaging 126~(3)
  (2004) 308--316.
\newblock \href {https://doi.org/10.1115/1.1772411}
  {\path{doi:10.1115/1.1772411}}.

\bibitem{Dmitruk2020AluminumAccumulator}
A.~Dmitruk, K.~Naplocha, J.~Grzeda, J.~W. Kaczmar,
  \href{https://www.mdpi.com/1996-1944/13/2/415}{{Aluminum Inserts for
  Enhancing Heat Transfer in PCM Accumulator}}, Materials 13~(2) (2020) 415.
\newblock \href {https://doi.org/10.3390/ma13020415}
  {\path{doi:10.3390/ma13020415}}.

\bibitem{Ruiz2017InvestigationMaterials}
J.~Ruiz, Y.~Ganatra, A.~Bruce, J.~Howarter, A.~M. Marconnet,
  \href{http://ieeexplore.ieee.org/document/7992499/}{{Investigation of
  aluminum foams and graphite fillers for improving the thermal conductivity of
  paraffin wax-based phase change materials}}, in: 2017 16th IEEE Intersociety
  Conference on Thermal and Thermomechanical Phenomena in Electronic Systems
  (ITherm), IEEE, 2017, pp. 384--389.
\newblock \href {https://doi.org/10.1109/ITHERM.2017.7992499}
  {\path{doi:10.1109/ITHERM.2017.7992499}}.

\bibitem{Soupremanien2016IntegrationElectronics}
U.~Soupremanien, H.~Szambolics, S.~Quenard, P.~Bouchut, M.~Roumanie,
  R.~Bottazzini, N.~Dunoyer,
  \href{http://ieeexplore.ieee.org/document/7517539/}{{Integration of metallic
  phase change material in power electronics}}, in: 2016 15th IEEE Intersociety
  Conference on Thermal and Thermomechanical Phenomena in Electronic Systems
  (ITherm), IEEE, 2016, pp. 125--133.
\newblock \href {https://doi.org/10.1109/ITHERM.2016.7517539}
  {\path{doi:10.1109/ITHERM.2016.7517539}}.

\bibitem{Bonner2011DieDevices}
R.~W. Bonner, T.~Desai, F.~Gao, X.~Tang, T.~Palacios, S.~Shin, M.~Kaviany,
  \href{https://ieeexplore.ieee.org/document/5767199/authors#authors}{{Die
  level thermal storage for improved cooling of pulsed devices}}, in: Annual
  IEEE Semiconductor Thermal Measurement and Management Symposium, IEEE, 2011,
  pp. 193--198.
\newblock \href {https://doi.org/10.1109/STHERM.2011.5767199}
  {\path{doi:10.1109/STHERM.2011.5767199}}.

\bibitem{Green2012DynamicCoolers}
C.~E. Green, A.~G. Fedorov, Y.~K. Joshi,
  \href{http://ieeexplore.ieee.org/lpdocs/epic03/wrapper.htm?arnumber=6231516}{{Dynamic
  thermal management of high heat flux devices using embedded solid-liquid
  phase change materials and solid state coolers}}, in: 13th InterSociety
  Conference on Thermal and Thermomechanical Phenomena in Electronic Systems,
  IEEE, 2012, pp. 853--862.
\newblock \href {https://doi.org/10.1109/ITHERM.2012.6231516}
  {\path{doi:10.1109/ITHERM.2012.6231516}}.

\bibitem{Shao2014On-chipSprinting}
L.~Shao, A.~Raghavan, L.~Emurian, M.~C. Papaefthymiou, T.~F. Wenisch, M.~M.
  Martin, K.~P. Pipe,
  \href{https://ieeexplore.ieee.org/document/6892211/authors#authors}{{On-chip
  phase change heat sinks designed for computational sprinting}}, in: Annual
  IEEE Semiconductor Thermal Measurement and Management Symposium, IEEE, 2014,
  pp. 29--34.
\newblock \href {https://doi.org/10.1109/SEMI-THERM.2014.6892211}
  {\path{doi:10.1109/SEMI-THERM.2014.6892211}}.

\bibitem{Gurrum2002ThermalMaterials}
S.~P. Gurrum, Y.~K. Joshi, J.~Kim,
  \href{https://www.tandfonline.com/doi/full/10.1080/10407780290059800}{{THERMAL
  MANAGEMENT OF HIGH TEMPERATURE PULSED ELECTRONICS USING METALLIC PHASE CHANGE
  MATERIALS}}, Numerical Heat Transfer, Part A: Applications 42~(8) (2002)
  777--790.
\newblock \href {https://doi.org/10.1080/10407780290059800}
  {\path{doi:10.1080/10407780290059800}}.

\bibitem{Yang2018AdvancesManagement}
X.-H. Yang, J.~Liu, \href{https://doi.org/10.1016/bs.aiht.2018.07.002
  https://linkinghub.elsevier.com/retrieve/pii/S0065271718300030}{{Advances in
  Liquid Metal Science and Technology in Chip Cooling and Thermal Management}},
  in: Advances in Heat Transfer, Vol.~50, Elsevier Ltd, 2018, pp. 187--300.
\newblock \href {https://doi.org/10.1016/bs.aiht.2018.07.002}
  {\path{doi:10.1016/bs.aiht.2018.07.002}}.

\bibitem{Shamberger2020ReviewApplications}
P.~J. Shamberger, N.~M. Bruno,
  \href{https://doi.org/10.1016/j.apenergy.2019.113955
  https://linkinghub.elsevier.com/retrieve/pii/S0306261919316423}{{Review of
  metallic phase change materials for high heat flux transient thermal
  management applications}}, Applied Energy 258~(September 2019) (2020) 113955.
\newblock \href {https://doi.org/10.1016/j.apenergy.2019.113955}
  {\path{doi:10.1016/j.apenergy.2019.113955}}.

\bibitem{Gonzalez-Nino2018ExperimentalMitigation}
D.~Gonzalez-Nino, L.~M. Boteler, D.~Ibitayo, N.~R. Jankowski, D.~Urciuoli,
  I.~M. Kierzewski, P.~O. Quintero,
  \href{https://doi.org/10.1016/j.ijheatmasstransfer.2017.09.039
  https://linkinghub.elsevier.com/retrieve/pii/S0017931017317106}{{Experimental
  evaluation of metallic phase change materials for thermal transient
  mitigation}}, International Journal of Heat and Mass Transfer 116 (2018)
  512--519.
\newblock \href {https://doi.org/10.1016/j.ijheatmasstransfer.2017.09.039}
  {\path{doi:10.1016/j.ijheatmasstransfer.2017.09.039}}.

\bibitem{Liu2018Data-drivenResults}
Y.~Liu, N.~Dinh, Y.~Sato, B.~Niceno,
  \href{https://linkinghub.elsevier.com/retrieve/pii/S1359431118329065}{{Data-driven
  modeling for boiling heat transfer: Using deep neural networks and
  high-fidelity simulation results}}, Applied Thermal Engineering 144 (2018)
  305--320.
\newblock \href {https://doi.org/10.1016/j.applthermaleng.2018.08.041}
  {\path{doi:10.1016/j.applthermaleng.2018.08.041}}.

\bibitem{Kim2020PredictionNetworks}
J.~Kim, C.~Lee,
  \href{https://www.cambridge.org/core/product/identifier/S0022112019008140/type/journal_article}{{Prediction
  of turbulent heat transfer using convolutional neural networks}}, Journal of
  Fluid Mechanics 882 (2020) A18.
\newblock \href {https://doi.org/10.1017/jfm.2019.814}
  {\path{doi:10.1017/jfm.2019.814}}.

\bibitem{Athavale2018ArtificialCenters}
J.~Athavale, Y.~Joshi, M.~Yoda,
  \href{https://ieeexplore.ieee.org/document/8419607/}{{Artificial Neural
  Network Based Prediction of Temperature and Flow Profile in Data Centers}},
  in: 2018 17th IEEE Intersociety Conference on Thermal and Thermomechanical
  Phenomena in Electronic Systems (ITherm), IEEE, 2018, pp. 871--880.
\newblock \href {https://doi.org/10.1109/ITHERM.2018.8419607}
  {\path{doi:10.1109/ITHERM.2018.8419607}}.

\bibitem{Song2011MultivariateNetworks}
Z.~Song, B.~T. Murray, B.~Sammakia,
  \href{https://asmedigitalcollection.asme.org/InterPACK/proceedings/InterPACK2011/44625/595/351272}{{Multivariate
  Prediction of Airflow and Temperature Distributions Using Artificial Neural
  Networks}}, in: ASME 2011 Pacific Rim Technical Conference and Exhibition on
  Packaging and Integration of Electronic and Photonic Systems, MEMS and NEMS:
  Volume 2, Vol.~2, ASMEDC, Portland, 2011, pp. 595--604.
\newblock \href {https://doi.org/10.1115/IPACK2011-52167}
  {\path{doi:10.1115/IPACK2011-52167}}.

\bibitem{Shrivastava2007DataNetwork}
S.~K. Shrivastava, J.~W. VanGilder, B.~G. Sammakia,
  \href{https://asmedigitalcollection.asme.org/InterPACK/proceedings/InterPACK2007/42770/765/324391}{{Data
  Center Cooling Prediction Using Artificial Neural Network}}, in: ASME 2007
  InterPACK Conference, Volume 1, Vol.~1, ASMEDC, 2007, pp. 765--771.
\newblock \href {https://doi.org/10.1115/IPACK2007-33432}
  {\path{doi:10.1115/IPACK2007-33432}}.

\bibitem{Gao2014MachineOptimization}
J.~Gao, R.~Jamidar, {Machine Learning Applications for Data Center
  Optimization}, Google White Paper (2014) 1--13.

\bibitem{Shi2019GeometryChannel}
X.~Shi, S.~Li, Y.~Mu, B.~Yin,
  \href{https://doi.org/10.1016/j.icheatmasstransfer.2019.03.009
  https://linkinghub.elsevier.com/retrieve/pii/S0735193319300685}{{Geometry
  parameters optimization for a microchannel heat sink with secondary flow
  channel}}, International Communications in Heat and Mass Transfer 104~(March)
  (2019) 89--100.
\newblock \href {https://doi.org/10.1016/j.icheatmasstransfer.2019.03.009}
  {\path{doi:10.1016/j.icheatmasstransfer.2019.03.009}}.

\bibitem{Li2020HeatOptimization}
P.~Li, D.~Guo, X.~Huang,
  \href{https://doi.org/10.1016/j.applthermaleng.2020.115060
  https://linkinghub.elsevier.com/retrieve/pii/S1359431119365548}{{Heat
  transfer enhancement in microchannel heat sinks with dual split-cylinder and
  its intelligent algorithm based fast optimization}}, Applied Thermal
  Engineering 171~(September 2019) (2020) 115060.
\newblock \href {https://doi.org/10.1016/j.applthermaleng.2020.115060}
  {\path{doi:10.1016/j.applthermaleng.2020.115060}}.

\bibitem{Deckard2019ConvergencePackaging}
M.~Deckard, P.~Shamberger, M.~Fish, M.~Berman, J.~Wang, L.~Boteler,
  \href{https://ieeexplore.ieee.org/document/8757334/}{{Convergence and
  Validation in ParaPower: A Design Tool for Phase Change Materials in
  Electronics Packaging}}, in: 2019 18th IEEE Intersociety Conference on
  Thermal and Thermomechanical Phenomena in Electronic Systems (ITherm), Vol.
  2019-May, IEEE, 2019, pp. 878--885.
\newblock \href {https://doi.org/10.1109/ITHERM.2019.8757334}
  {\path{doi:10.1109/ITHERM.2019.8757334}}.

\bibitem{Border2019IntegratingAnalysis}
L.~M. Border, S.~M. Miner, M.~Fish, M.~Berman, {Integrating heat sinks into a
  3D co-design network model for quick parametric analysis}, InterSociety
  Conference on Thermal and Thermomechanical Phenomena in Electronic Systems,
  ITHERM 2019-May (2019) 518--524.
\newblock \href {https://doi.org/10.1109/ITHERM.2019.08757416}
  {\path{doi:10.1109/ITHERM.2019.08757416}}.

\bibitem{BotelerMultiplePackaging}
L.~Boteler,
  \href{https://d2vrpothuwvesb.cloudfront.net/uploads/techops/fe1193cf-aa80-4708-b52a-762d2db1c00e/attachments/FY17-OCOH-ScMvr-33-Multiple-Dom_E7PFqYn.pdf}{{Multiple
  Domain Optimized Power Packaging}}.

\bibitem{CollierS.Miers2019THERMALMATERIALS}
{Collier S. Miers}, {THERMAL METROLOGY FOR WASTE HEAT SYSTEMS: THERMOELECTRICS
  TO PHASE CHANGE MATERIALS}, Ph.D. thesis, Purdue University (2019).

\bibitem{MatWebIndiumAlloy}
{MatWeb},
  \href{http://www.matweb.com/search/datasheet.aspx?matguid=cddd7436bd7a4003bd1eadbcaae7c351}{{Indium
  Corp. Indalloy 117 Bi-Pb-In-Sn-Cd Fusible Alloy}}.

\bibitem{Fukuoka1990NewAlloys}
Y.~Fukuoka, M.~Ishizuka,
  \href{https://iopscience.iop.org/article/10.1143/JJAP.29.1377}{{New Package
  Cooling Technology Using Low Melting Point Alloys}}, Japanese Journal of
  Applied Physics 29~(Part 1, No. 7) (1990) 1377--1384.
\newblock \href {https://doi.org/10.1143/JJAP.29.1377}
  {\path{doi:10.1143/JJAP.29.1377}}.

\bibitem{Baez2020MetallicCycles}
R.~B{\'{a}}ez, L.~E. Gonz{\'{a}}lez, M.~X. de~Jes{\'{u}}s-L{\'{o}}pez, P.~O.
  Quintero, L.~M. Boteler,
  \href{https://asmedigitalcollection.asme.org/electronicpackaging/article/doi/10.1115/1.4047063/1083333/Metallic-Phase-Change-Materials-Microstructural}{{Metallic
  Phase Change Material's Microstructural Stability Under Repetitive
  Melting/Solidification Cycles}}, Journal of Electronic Packaging 142~(3)
  (2020) 1--8.
\newblock \href {https://doi.org/10.1115/1.4047063}
  {\path{doi:10.1115/1.4047063}}.

\bibitem{Yang2017ExperimentalFins}
X.-H. Yang, S.-C. Tan, Y.-J. Ding, L.~Wang, J.~Liu, Y.-X. Zhou,
  \href{https://linkinghub.elsevier.com/retrieve/pii/S0735193317301707}{{Experimental
  and numerical investigation of low melting point metal based PCM heat sink
  with internal fins}}, International Communications in Heat and Mass Transfer
  87 (2017) 118--124.
\newblock \href {https://doi.org/10.1016/j.icheatmasstransfer.2017.07.001}
  {\path{doi:10.1016/j.icheatmasstransfer.2017.07.001}}.

\bibitem{PureTempPureTempSheet}
{PureTemp}, \href{http://www.puretemp.com/stories/puretemp-60-tds}{{PureTemp 60
  technical data sheet}}.

\bibitem{MathWorksParticleAlgorithm}
{MathWorks},
  \href{https://www.mathworks.com/help/gads/particle-swarm-optimization-algorithm.html}{{Particle
  Swarm Optimization Algorithm}}.

\bibitem{MathWorksHowWorks}
{MathWorks},
  \href{https://www.mathworks.com/help/gads/how-the-genetic-algorithm-works.html}{{How
  the genetic algorithm works}}.

\bibitem{MathWorksChooseFunction}
{MathWorks},
  \href{https://www.mathworks.com/help/deeplearning/ug/choose-a-multilayer-neural-network-training-function.html}{{Choose
  a Multilayer Neural Network Training Function}}.

\end{thebibliography}





\end{document}